\documentclass[12pt]{article} 
\usepackage{amsmath}
\usepackage{amsfonts}
\usepackage{dsfont}
\usepackage{axodraw}
\usepackage{subfigure}
\usepackage{rotating}

\def\pa{\partial}

\def\be{\begin{equation}}
\def\ee{\end{equation}}
\newcommand{\bea}{\begin{eqnarray}}
\newcommand{\eea}{\end{eqnarray}}
\newcommand{\col}{~,}
\newcommand{\pnt}{~.}
\newlength{\diameter}
\newcommand{\olett}[1]{
  \ensuremath{
  \text{$\settowidth{\diameter}{$\bigcirc$}%
  \bigcirc%
  \hspace{-0.5\diameter}%
  \makebox[0pt][c]{$\scriptstyle{#1}$}%
  \hspace{0.5\diameter}$}}}    
\newlength{\neglength}
\newcommand{\negphantom}[1]{\text{\settowidth{\neglength}{$#1$}$
                             \hspace{-\neglength}$}}         
\makeatletter
\newcommand{\redefinelabel}[1]{
  \def\@currentlabel{#1}}
\makeatother
\DeclareMathOperator{\tr}{tr}
\DeclareMathOperator{\sgn}{sgn}
\begin {document}
\begin{titlepage}
May 2002 \\
\begin{flushright}
HU Berlin-EP-02/21\\ 
\end{flushright}
\mbox{ }  \hfill hep-th/0205286
\vspace{5ex}
\Large
\begin {center}     
{\bf Some remarks on Feynman rules for\\ non-commutative gauge theories based
on \\groups $G\neq U(N)$}
\end {center}
\large
\vspace{1ex}
\begin{center}
Harald Dorn and Christoph Sieg \footnote{dorn@physik.hu-berlin.de, csieg@physik.hu-berlin.de}
\end{center}
\begin{center}
Humboldt--Universit\"at zu Berlin, Institut f\"ur Physik\\
Invalidenstra\ss e 110, D-10115 Berlin\\[2mm]  
\end{center}
\vspace{4ex}
\rm
\begin{center}
{\bf Abstract}
\end{center} 
\normalsize 

We study for subgroups $G\subseteq U(N)$ partial summations of the
$\theta$-expanded perturbation theory. On diagrammatic level  a summation 
procedure is established, which in the $U(N)$ case delivers the full
star-product induced rules. Thereby we uncover a cancellation mechanism 
between certain diagrams, which is crucial in the $U(N)$ case, but set
out of work for $G\subset U(N)$. In addition, an explicit 
proof is given that for $G\subset U(N),~G\neq U(M),~M<N$ there is no
partial summation of the $\theta $-expanded rules resulting in new
Feynman rules using the $U(N)$ star-product vertices and  
besides suitable modified propagators at most a $finite$ number of additional
building blocks. Finally, we show that certain $SO(N)$
Feynman rules conjectured in the literature cannot be derived from the
enveloping algebra approach. 
\vfill
\end{titlepage} 
\section{Introduction}

In recent years there has been a lot of interest in gauge theories
on non-commutative spaces and in particular in their relation to string 
theory, see \cite{dn,sz} for reviews. 

The gauge theory based on the gauge transformation
\footnote{We consider only the non-commutative version of Minkowski
space $[x^{\mu},x^{\nu}]=i\theta ^{\mu\nu}$ with constant $\theta ^{\mu\nu}$
and use the Moyal $\star $-product formulation.
To avoid problems with unitarity \cite{gm, frede} we restrict ourselves
to the case of space non-commutativity.}
\be
\delta _{\Lambda}A_{\mu}~=~\partial _{\mu}\Lambda ~-~iA_{\mu}\star \Lambda
~+~i\Lambda \star A_{\mu}
\label{1}
\ee
and the action
\be
S[A]~=~-\frac{1}{2g^2}\int dx ~\tr (F_{\mu\nu}\star F^{\mu\nu})\col
\label{2}
\ee
with
\be
F_{\mu\nu}~=~\partial _{\mu}A_{\nu}-\partial _{\nu}A_{\mu}-i[A_{\mu},A_{\nu}]_{\star}\col
\label{3}
\ee
is classically consistent for the gauge group $U(N)$ only \cite{mssw}.
This observation is based on the fact that the $\star$-commutator
of two Lie algebra valued quantities is again in the Lie algebra only
if the anticommutator of two generators is in the Lie algebra. 

On the string theory side one gets non-commutative gauge field
theories in a certain infinite tension limit in the presence of
a constant Neveu-Schwarz $B$-field \cite{sw}. Then, e.g. for the case of
$SO(N)$, the problem reappears as an obstruction for the implementation
of a $B$-field background for non-oriented strings. Attempts to
overcome this obstruction have been made in \cite{bsst,bsv}. A
discussion on the level of factorization properties of string tree
amplitudes can be found in \cite{ad}. 

In spite of the obstruction preventing, for gauge groups $G$ other
than $U(N)$, Lie algebra
valued gauge fields $A_{\mu}$, one nevertheless
can consistently construct non-commutative gauge field theories for all
gauge groups within the enveloping algebra approach \cite{wess}.
For subgroups of $U(N)$ this construction is equivalent to the following 
set up: First one expresses the non-commutative $U(N)$ gauge field and
gauge transformation $\Lambda$ via the Seiberg-Witten map \cite{sw}
\bea
A_{\mu}&=&a_{\mu}-\frac{1}{4}~\theta ^{\alpha\beta}\{a_{\alpha},\pa _{\beta}a_{\mu}+f_{\beta\mu}\}~+~O(\theta ^2)\col\nonumber\\
\Lambda &=&\lambda +\frac{1}{4}~\theta ^{\alpha\beta}\{\pa _{\alpha}\lambda ,a_{\beta}\}
~+~O(\theta ^2)
\label{4}
\eea 
in terms of a commutative $U(N)$ gauge field $a_{\mu}$ and gauge transformation
$\lambda$, respectively. After that both $a_{\mu}$ and $\lambda$ are
constrained to take values in the Lie algebra of the subgroup $G$
under discussion. 
In slightly different words the non-commutative version of the $G$ gauge
theory is defined by (\ref{1})-(\ref{3}) $and$ (\ref{4}) {\it as well
  as} the constraint $a,~\lambda ~\in ~ G$. 

Inserting (\ref{4}) into (\ref{2}) and expanding the $\star$-product
 in powers of $\theta $ yields the following action for $a_{\mu}$ 
\be
s[a]~:=~S[A[a]]~=~-\frac{1}{2g^2}\int dx~\tr(f_{\mu\nu}f^{\mu\nu})~+~\dots \col
\label{5}
\ee
where the dots stand for an infinite series of terms containing higher
derivatives and powers of $\theta$. After gauge fixing one immediately
can read off Feynman rules for the field $a_{\mu}$. Besides the standard
propagators and vertices for $a_{\mu}$ and the Faddeev-Popov ghosts
one has an infinite set of additional vertices with an increasing
number of legs, derivatives and powers of $\theta $. 
For our further discussion it is useful to stress that all these
vertices are generated by $\theta $-expansion of the whole action, i.e.
both the non-commutative kinetic and interaction term.
In the following we call this kind of perturbation theory the
$\theta $-{\it expanded perturbation theory} for the non-commutative
$G$-gauge theory. It has been extensively studied e.g. in
\cite{wien1,wien2,wess2}.\\ 

On the other side for the $ U(N)$ case it is straightforward to get directly
from (\ref{1})-(\ref{3}) (after gauge fixing) Feynman rules in terms of
$A_{\mu}$ and non-commutative Faddeev-Popov ghosts, see e.g. \cite{a,bs}.
The rules look very similar to that of standard (commutative) Yang-Mills
theory. Modifications are due to the presence of anticommutators of Lie
algebra generators, and the $\star $-product generates additional momentum
dependent exponential factors in the vertices. These factors are responsible
for the UV/IR effect \cite{msv}. This effect received a lot of attention
in particular with respect to its stringy origin and its implications for the
renormalization program. However, the UV/IR effect is not manifest in
$\theta$-expanded  perturbation theory for $U(N)$.

For $G\neq U(N)$, besides a conjecture for $SO(N)$ in \cite{bs2},
Feynman rules in terms of the full non-commutative $A_{\mu}$ are not known. 
Our goal in this paper is to get information on these rules
by studying some issues of partial summing the known $\theta$-expanded rules.
Such rules would allow to study UV/IR mixing similar to the $U(N)$ case.

One could also try to apply the full machinery of constrained quantization
to the combined problem of gauge fixing $and$ constraining to
gauge fields and gauge transformations whose inverse Seiberg-Witten map
is in $G$. However, as long as the SW map is available as
a power expansion in $\theta$ only, there seems to be little hope
to reach directly along this line Feynman rules with a finite number
of building blocks.\footnote{The $\theta$-expanded rules contain an
infinite number of vertices. It seems to us of little use to replace these
rules by another set of rules with an infinite number of vertices.
Therefore, incited by the $U(N)$ example, we are after rules with a
finite number of building blocks.} The situation could be different
for the $SU(N)$ case, where an alternative constraint has been
proposed in \cite{cd}.\\ 

The paper is organized as follows. In section~\ref{secgenframe} 
we formulate our questions
in precise technical terms. The original non-commutative interactions
are kept as suppliers of at least part of the vertices of our wanted Feynman
rules. The $\theta $-expanded perturbation theory is summed with
respect to the vertices generated by the expansion of the kinetic term
only.  
 
In section~\ref{secperttheo} we study this program in parallel for
both $U(N)$ and $G\subset U(N)$. Since the outcome for the $U(N)$ case
is a priori known, this case can serve as some check of the
diagrammatic analysis. Indeed we will find a cancellation mechanism
which guarantees the known result. On the other hand, 
this cancellation mechanism breaks down for subgroups $G\subset U(N)$
which are not equal to some $U(M),~M<N$. This already gives strong
arguments for the nonexistence of Feynman rules based on the original
non-commutative $U(N)$ three and four-point vertices supplemented by  
suitable modified propagators and by at most a finite number of
additional building blocks. These additional building blocks would be
related to the connected Green functions of the gauge field and the
ghosts obtained by summing the $\theta $-expansion of the kinetic term.  

However, strictly speaking, by itself the absence of a mechanism doing the job
in the $U(N)$ case does not exclude some other mechanism
enforcing the finiteness of the  number of additional building
blocks. Therefore, in section~\ref{secnptproof} we give an explicit
proof that the partial summation of $\theta $-expanded perturbation
theory, studied in the previous sections, yields non-vanishing Green
functions with an arbitrary large number of external points.

Section~\ref{secressource} is devoted to a modification of the partial
summation procedure designed to make contact with the $SO(N)$ rules
conjectured in ref. \cite{bs2}. At least for $SO(3)$ we will explicitly
prove that these rules cannot be derived from $\theta $-expanded
perturbation theory. 

Some more technical considerations related to
sections~\ref{secnptproof} and \ref{secressource} can be found
in appendices \ref{appnptproof} and \ref{appB}.
%%%%%%%%%%%%%%%%%%%%%%%%%%%%%%%%%%%%%%%
\section{The general framework}
\label{secgenframe}
We start with non-commutative $U(N)$ in Feynman gauge described in terms of the
gauge field $A_{\mu}$ and Faddeev-Popov ghosts $C$ and $\bar C$.
The Seiberg-Witten map for the ghost $C$ looks like that for the gauge
transformation in (\ref{4}) while the antighost is kept unchanged \cite{wien1},
i.e.
\bea
C&=&c~+~\frac{1}{4}~\theta ^{\alpha\beta}\{\pa _{\alpha}c ,a_{\beta}\}
~+~O(\theta ^2)\col\nonumber\\
\bar C&=&\bar c\pnt
\label{6}
\eea
Then we separate
\be
S[A,C,\bar C]~=~S_{\mbox{\scriptsize kin}}[A,C,\bar C]~+~S_I[A,C,\bar C]\col
\label{7}
\ee
with
\be
S_{\mbox{\scriptsize kin}}[A,C,\bar C]~=~-\frac{1}{g^2}\int dx~\tr ~\pa _{\mu}A_{\nu}
\pa ^{\mu}A^{\nu}~-~\int dx~\pa _{\mu}\bar C\pa ^{\mu} C\pnt
\label{8}
\ee
The generating functional for non-commutative $G$ Green functions is given by
\footnote{To keep the notation compact we make no distinction
between the group and its Lie algebra and understand space-time integration
and internal index summation in the source terms.}
\be
Z_{ G }[J,\bar{\eta},\eta ]~=~\int _{a,c,\bar c~\in G }DA~D\bar C~DC~e^{i(S[A,C,\bar C ]+
AJ+\bar{\eta}C+\bar C\eta )}\pnt
\label{9}
\ee
By the notation $\int _{a,c,\bar c~\in G }$ we indicate the
integration over $A,~C,~\bar C$ with the constraint that their image
under the inverse Seiberg-Witten map is in $G$, i.e. $a,~c,~\bar c~
\in  G$. For $U(N)$ the constraint is trivially solved by
$A_{\mu}=A^M_{\mu}T_M$ and free integration over $A^M_{\mu},~C^M,~\bar
C^M$. \\   

To explore the possibility
of non-commutative $G $ Feynman rules, which after some possible
projection work with the $U(N)$ vertices, we write (\ref{9}) using (\ref{7}) as
\be
Z_{ G}[J,\bar{\eta},\eta ]~=~e^{iS_I[\frac{\delta}{i\delta J},\frac{\delta}{i\delta\bar{\eta}},\frac{\delta}{i\delta \eta}]}~Z^{\mbox{\scriptsize kin}}_{G}
[J,\bar{\eta},\eta ]\col
\label{10}
\ee
with
\be   
Z^{\mbox{\scriptsize kin}}_{G}[J,\bar{\eta},\eta ]~=~\int _
{a,c,\bar c~\in G   }DA~D\bar C~DC~e^{i(S_{\mbox{\scriptsize kin}}[A,C,\bar C]+ AJ+\bar{\eta}C+\bar C\eta )}\pnt
\label{11}
\ee 
Denoting by ${\cal J}$ the functional determinant for changing the
integration variables from $A,~C,~\hat C$ to $a,~c,~\hat c$ we get 
\be   
Z^{\mbox{\scriptsize kin}}_{G}[J,\bar{\eta},\eta ]~=~\int _{a,c,\bar c~\in G   }Da~D\bar c~Dc~{\cal J}[a,c,\bar c ]~e^{i(S_{\mbox{\scriptsize kin}}[a,c,\bar c]+s_1[a,c,\bar c ]+A[a]J+\bar{\eta}C[c,a]+\bar c\eta )}\pnt
\label{12}
\ee
The new quantity $s_1[a,c,\bar c]$ appearing above is defined via (\ref{4}), (\ref{6}) and (\ref{8}) by
\be
S_{\mbox{\scriptsize kin}}[A[a],C[c,a],\bar c]~=~S_{\mbox{\scriptsize kin}}[a,
c,\bar c]~+~s_1[a,c,\bar c]\pnt
\label{13}
\ee
The logarithm of (\ref{12}) divided by $Z_G^{\mbox{\scriptsize
    kin}}[0,0,0]$ is the generating functional for the connected
Green functions of the composites $A,C,\bar C$ in the field theory
with elementary fields $a,c,\bar c$ interacting via $s_1-i\log {\cal
  J}$. Therefore it can be represented by
\be
\log (Z_G^{\mbox{\scriptsize kin}}[J,\bar{\eta},\eta ]/Z_G^{\mbox{\scriptsize kin}}[0])=
\sum _ni^n\int dx_1\dots dx_n~J(x_1)\dots J(x_n)
\langle A(x_1)\dots A(x_n)\rangle _{\mbox{\scriptsize kin} }~+~\dots \col
\label{13a}
\ee
where $\langle A(x_1)\dots A(x_n)\rangle _{\mbox{\scriptsize kin} }$
stands for the $n$-point connected Green function of $A$ in this field
theory. The dots at the end represent the corresponding ghost and
mixed ghost and gauge field terms. 

Neglecting ${\cal J}$ (for a justification see next section) these are just the
Green functions for the composites $A[a],C,\bar C$ obtained within
$\theta $-expanded perturbation theory by partial summation of all diagrams
built with vertices generated by the $\theta $-expansion of the non-commutative
kinetic term only.

In the $U(N)$-case $Z^{\mbox{\scriptsize kin}}_{U(N)}[J,\bar{\eta},\eta ]$
as given by (\ref{11}) is a trivial Gaussian integral. It is the generating
functional of Green functions for $A,~C,~\bar C$ treated as free fields.
Then in (\ref{13a}) only the two point functions $\langle
AA\rangle_{\mbox{\scriptsize kin} }$ and $\langle C\bar
C\rangle_{\mbox{\scriptsize kin} }$ are different from zero. In
addition they are equal to the free propagators. 
Inserting this into (\ref{10}) yields the non-commutative Feynman
rules for $U(N)$.

Starting with free fields and imposing a constraint, in the generic case,
generates an interacting theory. We want to decide what happens in our
case (\ref{11}) for $G\subset U(N)$.  
By some special circumstance it could be that only the connected two-point
functions are modified. Another less restrictive possibility would be
that connected $n$-point functions beyond some finite $n_0>2$
vanish. In both 
cases from (\ref{10}) we would get Feynman rules with a finite number
of building blocks.

For $U(N)$ the equivalent representation (\ref{12}) is due to
a simple field redefinition of a free theory. Therefore, looking at
the $n$-point functions of the, in terms of $a,c,\bar c$, composite
operators $A,C,\bar C$ (see (\ref{4}),(\ref{6})) the summation of the
perturbation theory with respect to $s_1[a,c,\bar c]-i\log {\cal J}$
must yield the free field result guaranteed by
(\ref{11}).\footnote{This is a manifestation of the equivalence
  theorem \cite{equiv}.}    

On the other side for $G\subset U(N)$ we cannot directly evaluate
(\ref{11}) and are forced to work with (\ref{12}). It will turn out
to be useful to study both $U(N)$ and $G\subset U(N)$ in parallel.
Since the result for $U(N)$ is a priori known, one has some checks
for the calculations within the $s_1$-perturbation theory. 

%%%%%%%%%%%%%%%%%%%%%%
\section{$s_1$-perturbation theory for $U(N)$ and $G\subset U(N)$}
\label{secperttheo}
In both cases our gauge fields take values in the Lie algebra of $U(N)$.
We write
\be
A_{\mu}~=~A_{\mu}^B~T_B
\label{14}
\ee
and use the following relations for the generators $T_A$ of the $U(N)$
Lie algebra
\be
[T_A,T_B]~=~if_{ABC}~T_C\col~~~~\{T_A,T_B\}~=~d_{ABC}~T_C\col~~~~\tr (T_AT_B)~=~
\frac{1}{2}\delta _{AB}\pnt
\label{15}
\ee
Then (\ref{4}) and (\ref{6}) imply
\bea
A_{\mu}^M&=&a_{\mu}^M~-~\frac{1}{2}\theta ^{\alpha\beta}a_{\alpha}^P\pa _{\beta
}a_{\mu}^Q~d_{MPQ}~+~\frac{1}{4}\theta ^{\alpha\beta}a_{\alpha}^P\pa _{\mu}
a_{\beta}^Q~d_{MPQ}\nonumber\\
&-&\frac{1}{4}\theta ^{\alpha\beta}a_{\alpha}^Pa_{\beta}^Qa_{\mu}^R~d_{MPS}f_{
SQR}~+~O(\theta ^2)
\label{16}
\eea
and
\bea
C^M&=&c^M~+~\frac{1}{4}\theta ^{\alpha\beta}\pa _{\alpha}c^Pa_{\beta }^Q~
d_{MPQ}~+~O(\theta ^2)\nonumber\\
\bar C^M&=&\bar c^M\pnt
\label{17}
\eea
In the case $G\subset U(N)$, $G\neq U(M)$, $M<N$~\footnote{In the
  following we sometimes implicitly understand that $G\subset U(N)$
  excludes $U(M)$ subgroups.} we indicate the {\it
  generators spanning the Lie algebra of $G$} with a {\it lower case}
Latin index and the {\it remaining} ones with a {\it primed lower
  case} Latin index. Upper case Latin indices run over all $U(N)$ generators.
Since $G$ is a subgroup and since $\{T_a,T_b\}$ is not in the Lie algebra
of $G$ we have
\be
f_{abc'}~=~0\col\forall ~a,b,c'~~~~\mbox{and}~~~d_{abc'}\neq 0~~~\mbox{for some}~a,b,c'
\pnt
\label{18}
\ee
As discussed in the previous section the non-commutative $G$ gauge
field theory is then defined by unconstrained functional integration
over $a_{\mu}^b,c^b,\bar c^b$ $and$
\be
a_{\mu}^{b'}~=~c^{b'}~=~\bar c^{b'}~=~0\pnt
\label{19}
\ee
In spite of (\ref{19}) via (\ref{16})-(\ref{18}) one has non-vanishing
$A_{\mu}^{b'}$ and $C^{b'}$.\\ 

We are interested in (\ref{12}), i.e. the Green functions of $A,~C,~\bar C$,
which are composites in terms of $a,~c,~\bar c$. For the diagrammatic
evaluation one gets from (\ref{16}) the $external$ vertices where all
momenta are directed to the interaction point and a slash denotes
derivative of the field at the corresponding leg. (We write down the
$\propto\theta^0$ and $\propto\theta^1$ contributions only. Momentum
conservation at all vertices is understood.) 
\newlength{\eqoff}
\settoheight{\eqoff}{\fbox{$=$}}
\setlength{\eqoff}{0.5\eqoff}
\addtolength{\eqoff}{-40pt}
\begin{equation}
\raisebox{\eqoff}{%
\begin{picture}(120,80)(0,0)\scriptsize
%\Boxc(60,40)(120,80)
\Vertex(40,40){1}\Text(30,40)[r]{$p,\mu,M$}
\Line(40,40)(80,40)\Text(90,40)[l]{$k,\alpha,A$}
\end{picture}}
=
\begin{cases}
\delta ^{\alpha}_{\mu}~\delta _{AM} & \text{for $M=m$}\\
0 & \text{for $M=m'$}
\end{cases}\col
\label{20}
\end{equation}
\be
\raisebox{\eqoff}{%
\begin{picture}(120,80)(0,0)\scriptsize
%\Boxc(60,40)(120,80)
\Vertex(40,40){1}\Text(30,40)[r]{$p,\mu,M$}
\Line(40,40)(74.641,60)\Text(84.641,60)[l]{$k_2,\beta,B$}
\Line(40,40)(74.641,20)\Text(84.641,20)[l]{$k_1,\alpha,A$}
\Line(43.08,35.335)(45.58,39.665)
\end{picture}}
~=~i\left ( \frac{1}{4}\theta ^{\beta\alpha}\delta _{\mu}^{\nu}
d_{MAB}~-~\frac{1}{2}\theta ^{\beta\nu}\delta _{\mu}^{\alpha}d_{MAB}\right )(k
_1)_{\nu}\col
\label{21}
\ee
\be
\raisebox{\eqoff}{%
\begin{picture}(120,80)(0,0)\scriptsize
%\Boxc(60,40)(120,80)
\Vertex(40,40){1}\Text(30,40)[r]{$p,\mu,M$}
\Line(40,40)(74.641,60)\Text(84.641,60)[l]{$k_3,\gamma,C$}
\Line(40,40)(80,40)\Text(90,40)[l]{$k_2,\beta,B$}
\Line(40,40)(74.641,20)\Text(84.641,20)[l]{$k_1,\alpha,A$}
\end{picture}}
~=~-\frac{1}{4}~\theta ^{\alpha\beta}d_{MAE}f_{EBC}\delta^\gamma_\mu\col
\label{22}
\ee
and
\begin{equation}
\settoheight{\eqoff}{\fbox{$=$}}
\setlength{\eqoff}{0.5\eqoff}
\addtolength{\eqoff}{-40pt}
\raisebox{\eqoff}{%
\begin{picture}(140,80)(0,0)\scriptsize
%\Boxc(70,40)(140,80)
\Vertex(40,40){1}\Text(30,40)[r]{$p,M$}
\DashArrowLine(80,40)(40,40){1}\Text(90,40)[l]{$k,\alpha,A$}
\end{picture}}
=
\raisebox{\eqoff}{%
\begin{picture}(120,80)(0,0)\scriptsize
%\Boxc(60,40)(120,80)
\Vertex(40,40){1}\Text(30,40)[r]{$p,M$}
\DashArrowLine(40,40)(80,40){1}\Text(90,40)[l]{$k,A$}
\end{picture}}
=
\begin{cases}
\delta _{AM} & \text{for $M=m$}\\
0 & \text{for $M=m'$}
\end{cases}
\col
\label{23}
\end{equation}
\be
\raisebox{\eqoff}{%
\begin{picture}(120,80)(0,0)\scriptsize
%\Boxc(60,40)(120,80)
\Vertex(40,40){1}\Text(30,40)[r]{$p,M$}
\Line(40,40)(74.641,60)\Text(84.641,60)[l]{$k_2,\beta,B$}
\DashArrowLine(74.641,20)(40,40){1}\Text(84.641,20)[l]{$k_1,A$}
\Line(43.08,35.335)(45.58,39.665)
\end{picture}}
~=~ \frac{i}{4}\theta ^{\nu\beta}d_{MAB}(k_1)_{\nu}\pnt
\label{24}
\ee
The insertion of (\ref{16}) and (\ref{17}) into (\ref{13}) yields
$s_1[a,c,\bar c]$ generating the {\it internal} vertices\\
\settoheight{\eqoff}{\fbox{$=$}}
\setlength{\eqoff}{0.5\eqoff}
\addtolength{\eqoff}{-40pt}
\be
\raisebox{\eqoff}{%
\begin{picture}(150,80)(0,0)\scriptsize
%\Boxc(75,40)(150,80)
\SetOffset(35,0)
\Line(0,40)(40,40)\Text(-5,40)[r]{$k_1,\alpha,A$}
\Line(32.5,37.5)(32.5,42.5)
\Line(35,37.5)(35,42.5)
\Vertex(40,40){1}
\Line(40,40)(74.641,60)\Text(84.641,60)[lb]{$k_3,\gamma,C$}
\Line(40,40)(74.641,20)\Text(84.641,20)[lt]{$k_2,\beta,B$}
\Line(43.08,35.335)(45.58,39.665)
\end{picture}}
~=~\frac{1}{g^2}\left (\frac{1}{4}\theta ^{\gamma\beta}g^{\nu\alpha}-\frac{1}{2}\theta ^{\gamma\nu}g^{\alpha\beta}\right )d_{ABC}k_1^2(k_2)_{\nu}
\col
\label{25}
\ee
\be
\raisebox{\eqoff}{%
\begin{picture}(155,80)(0,0)\scriptsize
%\Boxc(77.5,40)(155,80)
\SetOffset(35,0)
\Line(0,40)(40,40)\Text(-5,40)[r]{$k_1,\alpha,A$}
\Line(32.5,37.5)(32.5,42.5)
\Line(35,37.5)(35,42.5)
\Vertex(40,40){1}
\Line(40,40)(74.641,60)\Text(84.641,60)[lb]{$k_4,\delta,D$}
\Line(40,40)(80,40)\Text(90,40)[l]{$k_3,\gamma,C$}
\Line(40,40)(74.641,20)\Text(84.641,20)[lt]{$k_2,\beta,B$}
\end{picture}}
~=~\frac{i}{4g^2}\theta ^{\delta\gamma}d_{ADE}f_{ECB}g^{\alpha\beta} k_1^2\col
\label{26}
\ee
\be
\raisebox{\eqoff}{%
\begin{picture}(150,80)(0,0)\scriptsize
%\Boxc(75,40)(150,80)
\SetOffset(35,0)
\DashArrowLine(40,40)(0,40){1}\Text(-5,40)[r]{$k_1,A$}
\Line(32.5,37.5)(32.5,42.5)
\Line(35,37.5)(35,42.5)
\Vertex(40,40){1}
\DashArrowLine(74.641,60)(40,40){1}\Text(84.641,60)[lb]{$k_3,C$}
\Line(40,40)(74.641,20)\Text(84.641,20)[lt]{$k_2,\beta,B$}
\Line(43.08,44.6655)(45.58,40.335)
\end{picture}}
~=~-\frac{1}{4}\theta ^{\beta\nu}d_{ABC}(k_3)_{\nu}k_1^2\pnt
\label{27}
\ee
The double slash stems from the derivatives in (\ref{8}) after
partial integration and denotes the action of
$\Box=\partial_\mu\partial^\mu$ at the corresponding leg.

%%%%%%%%%%%%%%%
\begin{figure}
\begin{center}
\subfigure[]{%
\begin{picture}(30,30)(0,0)\scriptsize
%\Boxc(15,15)(30,30)
\Vertex(5,15){1}
\Vertex(25,15){1}
\Line(5,15)(25,15)
\end{picture}\label{one}}\qquad
\subfigure[]{%
\begin{picture}(30,30)(0,0)\scriptsize
%\Boxc(15,15)(30,30)
\Vertex(5,15){1}
\Vertex(25,15){1}
\CArc(15,9.2265)(11.547,30,150)
\Line(11.25,17.0615)(8.75,21.3915)
%\Vertex(10,19.2265){1}
\Line(18.75,17.0615)(21.25,21.3915)
%\Vertex(20,19.2265){1}
\CArc(15,20.7735)(11.547,-150,-30)
%\Vertex(10,10.7735){1}
%\Line(11.25,12.9385)(8.75,8.6085)
%\Vertex(20,10.7735){1}
%\Line(18.75,12.9385)(21.25,8.6085)
\end{picture}\label{two}}\qquad
\subfigure[]{%
\begin{picture}(50,30)(0,0)\scriptsize
%\Boxc(25,15)(50,30)
\SetOffset(20,0)
\Vertex(-15,15){1}
\Line(-15,15)(5,15)
\Line(0,12.5)(0,17.5)
\Line(-2.5,12.5)(-2.5,17.5)
\Vertex(5,15){1}
\Vertex(25,15){1}
\CArc(15,9.2265)(11.547,30,150)
\Line(11.25,17.0615)(8.75,21.3915)
%\Vertex(10,19.2265){1}
\Line(18.75,17.0615)(21.25,21.3915)
%\Vertex(20,19.2265){1}
\CArc(15,20.7735)(11.547,-150,-30)
%\Vertex(10,10.7735){1}
%\Line(11.25,12.9385)(8.75,8.6085)
%\Vertex(20,10.7735){1}
%\Line(18.75,12.9385)(21.25,8.6085)
\end{picture}\label{three}}\qquad
\subfigure[]{%
\begin{picture}(50,30)(0,0)\scriptsize
%\Boxc(25,15)(50,30)
\SetOffset(30,0)
\Vertex(15,15){1}
\Line(15,15)(-5,15)
\Line(0,12.5)(0,17.5)
\Line(2.5,12.5)(2.5,17.5)
\Vertex(-5,15){1}
\Vertex(-25,15){1}
\CArc(-15,9.2265)(11.547,30,150)
\Line(-11.25,17.0615)(-8.75,21.3915)
%\Vertex(-10,19.2265){1}
\Line(-18.75,17.0615)(-21.25,21.3915)
%\Vertex(-20,19.2265){1}
\CArc(-15,20.7735)(11.547,-150,-30)
%\Vertex(-10,10.7735){1}
%\Line(-11.25,12.9385)(-8.75,8.6085)
%\Vertex(-20,10.7735){1}
%\Line(-18.75,12.9385)(-21.25,8.6085)
\end{picture}\label{four}}\qquad
\subfigure[]{%
\begin{picture}(70,30)(0,0)\scriptsize
%\Boxc(35,15)(70,30)
\SetOffset(20,0)
\Vertex(-15,15){1}
\Line(-15,15)(5,15)
\Line(0,12.5)(0,17.5)
\Line(-2.5,12.5)(-2.5,17.5)
\Vertex(5,15){1}
\Vertex(25,15){1}
\CArc(15,9.2265)(11.547,30,150)
\Line(11.25,17.0615)(8.75,21.3915)
%\Vertex(10,19.2265){1}
\Line(18.75,17.0615)(21.25,21.3915)
%\Vertex(20,19.2265){1}
\CArc(15,20.7735)(11.547,-150,-30)
%\Vertex(10,10.7735){1}
%\Line(11.25,12.9385)(8.75,8.6085)
%\Vertex(20,10.7735){1}
%\Line(18.75,12.9385)(21.25,8.6085)
\Vertex(45,15){1}
\Line(25,15)(45,15)
\Line(30,12.5)(30,17.5)
\Line(32.5,12.5)(32.5,17.5)
\end{picture}\label{five}}\\
%
% second line of subfigures
%
\begin{picture}(30,30)(0,0)\scriptsize
%\Boxc(15,15)(30,30)
\end{picture}\qquad
\subfigure[]{%
\begin{picture}(30,30)(0,0)\scriptsize
%\Boxc(15,15)(30,30)
\Vertex(5,15){1}
\Vertex(25,15){1}
\CArc(15,9.2265)(11.547,30,150)
\Line(11.25,17.0615)(8.75,21.3915)
%\Vertex(10,19.2265){1}
%\Line(18.75,17.0615)(21.25,21.3915)
%\Vertex(20,19.2265){1}
\CArc(15,20.7735)(11.547,-150,-30)
%\Vertex(10,10.7735){1}
%\Line(11.25,12.9385)(8.75,8.6085)
%\Vertex(20,10.7735){1}
\Line(18.75,12.9385)(21.25,8.6085)
\end{picture}\label{six}}\qquad
\subfigure[]{%
\begin{picture}(50,30)(0,0)\scriptsize
%\Boxc(25,15)(50,30)
\SetOffset(20,0)
\Vertex(-15,15){1}
\Line(-15,15)(5,15)
\Line(0,12.5)(0,17.5)
\Line(-2.5,12.5)(-2.5,17.5)
\Vertex(5,15){1}
\Vertex(25,15){1}
\CArc(15,9.2265)(11.547,30,150)
\Line(11.25,17.0615)(8.75,21.3915)
%\Vertex(10,19.2265){1}
%\Line(18.75,17.0615)(21.25,21.3915)
%\Vertex(20,19.2265){1}
\CArc(15,20.7735)(11.547,-150,-30)
%\Vertex(10,10.7735){1}
%\Line(11.25,12.9385)(8.75,8.6085)
%\Vertex(20,10.7735){1}
\Line(18.75,12.9385)(21.25,8.6085)
\end{picture}\label{seven}}\qquad
\subfigure[]{%
\begin{picture}(50,30)(0,0)\scriptsize
%\Boxc(25,15)(50,30)
\SetOffset(30,0)
\Vertex(15,15){1}
\Line(15,15)(-5,15)
\Line(0,12.5)(0,17.5)
\Line(2.5,12.5)(2.5,17.5)
\Vertex(-5,15){1}
\Vertex(-25,15){1}
\CArc(-15,9.2265)(11.547,30,150)
%\Line(-11.25,17.0615)(-8.75,21.3915)
%\Vertex(-10,19.2265){1}
\Line(-18.75,17.0615)(-21.25,21.3915)
%\Vertex(-20,19.2265){1}
\CArc(-15,20.7735)(11.547,-150,-30)
%\Vertex(-10,10.7735){1}
\Line(-11.25,12.9385)(-8.75,8.6085)
%\Vertex(-20,10.7735){1}
%\Line(-18.75,12.9385)(-21.25,8.6085)
\end{picture}\label{eight}}\qquad
\subfigure[]{%
\begin{picture}(70,30)(0,0)\scriptsize
%\Boxc(35,15)(70,30)
\SetOffset(20,0)
\Vertex(-15,15){1}
\Line(-15,15)(5,15)
\Line(0,12.5)(0,17.5)
\Line(-2.5,12.5)(-2.5,17.5)
\Vertex(5,15){1}
\Vertex(25,15){1}
\CArc(15,9.2265)(11.547,30,150)
\Line(11.25,17.0615)(8.75,21.3915)
%\Vertex(10,19.2265){1}
%\Line(18.75,17.0615)(21.25,21.3915)
%\Vertex(20,19.2265){1}
\CArc(15,20.7735)(11.547,-150,-30)
%\Vertex(10,10.7735){1}
%\Line(11.25,12.9385)(8.75,8.6085)
%\Vertex(20,10.7735){1}
\Line(18.75,12.9385)(21.25,8.6085)
\Vertex(45,15){1}
\Line(25,15)(45,15)
\Line(30,12.5)(30,17.5)
\Line(32.5,12.5)(32.5,17.5)
\end{picture}\label{nine}}\\
%
% third line of subfigures
%
\begin{picture}(30,30)(0,0)\scriptsize
%\Boxc(15,15)(30,30)
\end{picture}\qquad
\subfigure[]{%
\begin{picture}(30,30)(0,0)\scriptsize
%\Boxc(15,15)(30,30)
\Vertex(5,15){1}
\Vertex(25,15){1}
\CArc(15,9.2265)(11.547,30,150)
%\Line(11.25,17.0615)(8.75,21.3915)
%\Vertex(10,19.2265){1}
%\Line(18.75,17.0615)(21.25,21.3915)
%\Vertex(20,19.2265){1}
\CArc(15,20.7735)(11.547,-150,-30)
%\Vertex(10,10.7735){1}
%\Line(11.25,12.9385)(8.75,8.6085)
%\Vertex(20,10.7735){1}
%\Line(18.75,12.9385)(21.25,8.6085)
\Line(5,15)(25,15)
\end{picture}\label{ten}}\qquad
\subfigure[]{%
\begin{picture}(50,30)(0,0)\scriptsize
%\Boxc(25,15)(50,30)
\SetOffset(20,0)
\Vertex(-15,15){1}
\Line(-15,15)(5,15)
\Line(0,12.5)(0,17.5)
\Line(-2.5,12.5)(-2.5,17.5)
\Vertex(5,15){1}
\Vertex(25,15){1}
\CArc(15,9.2265)(11.547,30,150)
%\Line(11.25,17.0615)(8.75,21.3915)
%\Vertex(10,19.2265){1}
%\Line(18.75,17.0615)(21.25,21.3915)
%\Vertex(20,19.2265){1}
\CArc(15,20.7735)(11.547,-150,-30)
%\Vertex(10,10.7735){1}
%\Line(11.25,12.9385)(8.75,8.6085)
%\Vertex(20,10.7735){1}
%\Line(18.75,12.9385)(21.25,8.6085)
\Line(5,15)(25,15)
\end{picture}\label{eleven}}\qquad
\subfigure[]{%
\begin{picture}(50,30)(0,0)\scriptsize
%\Boxc(25,15)(50,30)
\SetOffset(30,0)
\Vertex(15,15){1}
\Line(15,15)(-5,15)
\Line(0,12.5)(0,17.5)
\Line(2.5,12.5)(2.5,17.5)
\Vertex(-5,15){1}
\Vertex(-25,15){1}
\CArc(-15,9.2265)(11.547,30,150)
%\Line(-11.25,17.0615)(-8.75,21.3915)
%\Vertex(-10,19.2265){1}
%\Line(-18.75,17.0615)(-21.25,21.3915)
%\Vertex(-20,19.2265){1}
\CArc(-15,20.7735)(11.547,-150,-30)
%\Vertex(-10,10.7735){1}
%\Line(-11.25,12.9385)(-8.75,8.6085)
%\Vertex(-20,10.7735){1}
%\Line(-18.75,12.9385)(-21.25,8.6085)
\Line(-5,15)(-25,15)
\end{picture}\label{twelve}}\qquad
\subfigure[]{%
\begin{picture}(70,30)(0,0)\scriptsize
%\Boxc(35,15)(70,30)
\SetOffset(20,0)
\Vertex(-15,15){1}
\Line(-15,15)(5,15)
\Line(0,12.5)(0,17.5)
\Line(-2.5,12.5)(-2.5,17.5)
\Vertex(5,15){1}
\Vertex(25,15){1}
\CArc(15,9.2265)(11.547,30,150)
%\Line(11.25,17.0615)(8.75,21.3915)
%\Vertex(10,19.2265){1}
%\Line(18.75,17.0615)(21.25,21.3915)
%\Vertex(20,19.2265){1}
\CArc(15,20.7735)(11.547,-150,-30)
%\Vertex(10,10.7735){1}
%\Line(11.25,12.9385)(8.75,8.6085)
%\Vertex(20,10.7735){1}
%\Line(18.75,12.9385)(21.25,8.6085)
\Line(5,15)(25,15)
\Vertex(45,15){1}
\Line(25,15)(45,15)
\Line(30,12.5)(30,17.5)
\Line(32.5,12.5)(32.5,17.5)
\end{picture}\label{thirteen}}
\end{center}
\caption{Contributions to $\langle A_{\mu}^MA_{\nu}^N\rangle
  _{\mbox{\scriptsize kin}} $ up to order $\theta^2$} 
\label{propcorr}
\end{figure}
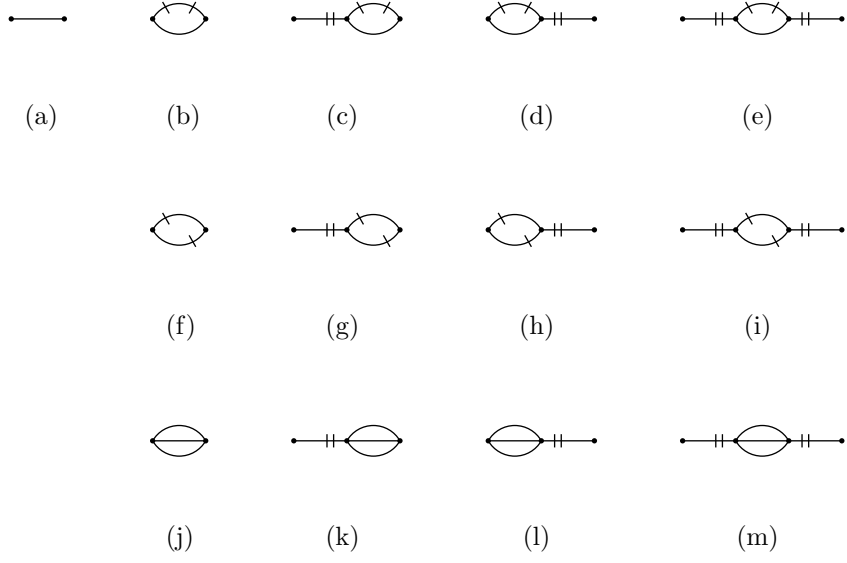

\settoheight{\eqoff}{\fbox{$=$}}
\setlength{\eqoff}{0.5\eqoff}
\addtolength{\eqoff}{-15pt}
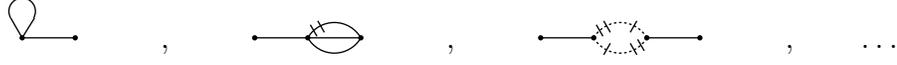
\begin{figure}
\begin{center}
\raisebox{\eqoff}{%
\begin{picture}(35,30)(0,0)\scriptsize
%\Boxc(15,15)(30,30)
\SetOffset(5,0)
\Vertex(5,15){1}
\Vertex(25,15){1}
\Line(5,15)(0.67,22.5)
\Line(5,15)(9.67,22.5)
\CArc(5,25)(5,-30,210)
\Line(5,15)(25,15)
\end{picture}}
$\qquad,\qquad$
\raisebox{\eqoff}{%
\begin{picture}(50,30)(0,0)\scriptsize
%\Boxc(25,15)(50,30)
\SetOffset(20,0)
\Vertex(-15,15){1}
\Line(-15,15)(5,15)
%\Line(0,12.5)(0,17.5)
%\Line(-2.5,12.5)(-2.5,17.5)
\Vertex(5,15){1}
\Vertex(25,15){1}
\CArc(15,9.2265)(11.547,30,150)
\Line(11.25,17.0615)(8.75,21.3915)
%\Vertex(10,19.2265){1}
\Line(8.4631,15.587)(5.9631,19.917)
%\Vertex(7.2131,17.752){1}
%\Line(18.75,17.0615)(21.25,21.3915)
%\Vertex(20,19.2265){1}
\CArc(15,20.7735)(11.547,-150,-30)
%\Vertex(10,10.7735){1}
%\Line(11.25,12.9385)(8.75,8.6085)
%\Vertex(20,10.7735){1}
%\Line(18.75,12.9385)(21.25,8.6085)
\Line(5,15)(25,15)
\end{picture}}
$\qquad,\qquad$
\raisebox{\eqoff}{%
\begin{picture}(70,30)(0,0)\scriptsize
%\Boxc(35,15)(70,30)
\SetOffset(20,0)
\Vertex(-15,15){1}
\Line(-15,15)(5,15)
%\Line(0,12.5)(0,17.5)
%\Line(-2.5,12.5)(-2.5,17.5)
\Vertex(5,15){1}
\Vertex(25,15){1}
\DashCArc(15,9.2265)(11.547,30,150){1}
\Line(11.25,17.0615)(8.75,21.3915)
%\Vertex(10,19.2265){1}
\Line(8.4631,15.587)(5.9631,19.917)
%\Vertex(7.2131,17.752){1}
\Line(18.75,17.0615)(21.25,21.3915)
%\Vertex(20,19.2265){1}
\DashCArc(15,20.7735)(11.547,-150,-30){1}
%\Vertex(10,10.7735){1}
\Line(11.25,12.9385)(8.75,8.6085)
%\Vertex(20,10.7735){1}
\Line(18.75,12.9385)(21.25,8.6085)
%\Vertex(7.2131,17.752){1}
\Line(21.5369,14.413)(24.0369,10.083)
\Vertex(45,15){1}
\Line(25,15)(45,15)
%\Line(30,12.5)(30,17.5)
%\Line(32.5,12.5)(32.5,17.5)
\end{picture}}
$\qquad,\qquad\dots$
\end{center}
\caption{Some additional vanishing contributions to $\langle
  A_{\mu}^MA_{\nu}^N\rangle _{\mbox{\scriptsize kin}} $ }
\label{detcontrib}
\end{figure}

The propagators are
\be
-ig^2~g_{\alpha\beta}\delta _{AB}\frac{1}{k^2}\col~~~~~~~~~
i\delta_{AB}\frac{1}{k^2} 
\label{28} 
\ee
for the commuting gauge field and ghosts, respectively.

Up to now we have not taken into account the functional determinant ${\cal J}$
in (\ref{12}). To simplify the analysis we use dimensional regularization.
Then this determinant is equal to one, and all diagrams containing
momentum integrals not depending on any external momentum or mass
parameter (tadpole type)
are zero, see e.g. \cite{zinn}. For other regularizations these
tadpole type diagrams just cancel the determinant contributions, at least
in the $U(N)$ case.\\

After these preparations we consider the 2-point function  $\langle
A_{\mu}^MA_{\nu}^N\rangle _{\mbox{\scriptsize kin}} $
within the perturbation theory with respect to $s_1$, see (\ref{12}). 
Fig.~\ref{propcorr} shows all diagrams
up to order $\theta ^2$ which do not vanish in dimensional regularization.
To give an impression, fig.~\ref{detcontrib} presents a part of the
remaining vanishing diagrams.

Let us first continue with the $U(N)$ case. Then a straightforward analysis
shows that the diagrams in fig.~\ref{two} - \ref{five} cancel among
each other. The same is true for fig.~\ref{six} - \ref{nine} and for
fig.~\ref{ten} - \ref{thirteen}.
 
The cancellation mechanism is quite general. Let us denote by
$M(k_1,\alpha ,A\vert k_2,\beta ,B)$ an arbitrary sub-diagram with two
marked legs, denoted by a shaded bubble in
fig.~\ref{cancelmech1}. Then the sum of the two diagrams in
fig.~\ref{cancelmech1} is equal to 
\bea
&&-\frac{ig^2}{p^2}g_{\mu\nu}\delta _{MN}~~\frac{1}{g^2}\left (\frac{1}{4}
\theta ^{\beta\alpha}g^{\lambda\nu}-\frac{1}{2}\theta ^{\beta\lambda}g^{\nu\alpha}\right )d_{NAB}~p^2(k_1)_{\lambda}~~M(k_1,\alpha ,A\vert k_2,\beta ,B)
\nonumber\\
&&+i\left (\frac{1}{4}\theta ^{\beta\alpha}\delta _{\mu}^{\lambda}-\frac{1}{2}
\theta ^{\beta\lambda}\delta ^{\alpha}_{\mu}\right )d_{MAB}(k_1)_{\lambda}
~~M(k_1,\alpha ,A\vert k_2,\beta ,B)~=~0\pnt
\label{29}
\eea
A similar general cancellation mechanism holds for diagrams of the
type shown in fig.~\ref{cancelmech2}. 

%%%%%%%%%%%%%%% 
\settoheight{\eqoff}{\fbox{$=$}}
\setlength{\eqoff}{0.5\eqoff}
\addtolength{\eqoff}{-40pt}
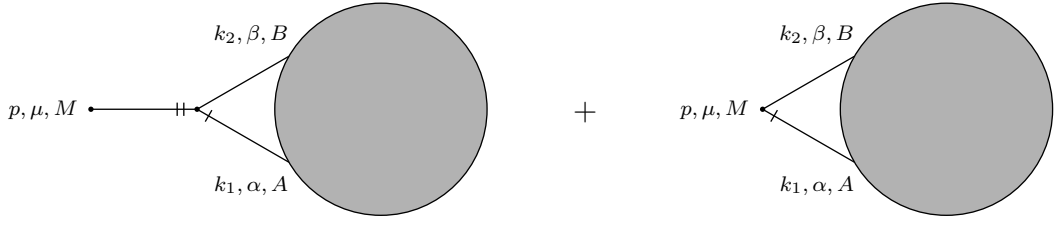
\begin{figure}
\begin{center}
\raisebox{\eqoff}{%
\begin{picture}(190,80)(0,0)\scriptsize
%\Boxc(95,40)(190,80)
\SetOffset(35,0)
\Line(0,40)(40,40)\Text(-5,40)[r]{$p,\mu,M$}
\Vertex(0,40){1}
\Line(32.5,37.5)(32.5,42.5)
\Line(35,37.5)(35,42.5)
\Vertex(40,40){1}
\Line(40,40)(74.641,60)\Text(74.641,65)[rb]{$k_2,\beta,B$}
\Line(40,40)(74.641,20)\Text(74.641,15)[rt]{$k_1,\alpha,A$}
\Line(43.08,35.335)(45.58,39.665)
\GCirc(109.282,40){40}{0.7}
\end{picture}}
$\qquad+\qquad$
\raisebox{\eqoff}{%
\begin{picture}(150,80)(0,0)\scriptsize
%\Boxc(75,40)(150,80)
\SetOffset(-5,0)
%\Line(0,40)(40,40)
%\Line(32.5,37.5)(32.5,42.5)
%\Line(35,37.5)(35,42.5)
\Vertex(40,40){1}\Text(35,40)[r]{$p,\mu,M$}
\Line(40,40)(74.641,60)\Text(74.641,65)[rb]{$k_2,\beta,B$}
\Line(40,40)(74.641,20)\Text(74.641,15)[rt]{$k_1,\alpha,A$}
\Line(43.08,35.335)(45.58,39.665)
\GCirc(109.282,40){40}{0.7}
\end{picture}}
\end{center}
\caption{Cancelling graphs from the $\propto a^2$ terms of the
  $\propto\theta$ terms of the SW map} 
\label{cancelmech1}
\end{figure}
\settoheight{\eqoff}{\fbox{$=$}}
\setlength{\eqoff}{0.5\eqoff}
\addtolength{\eqoff}{-40pt}
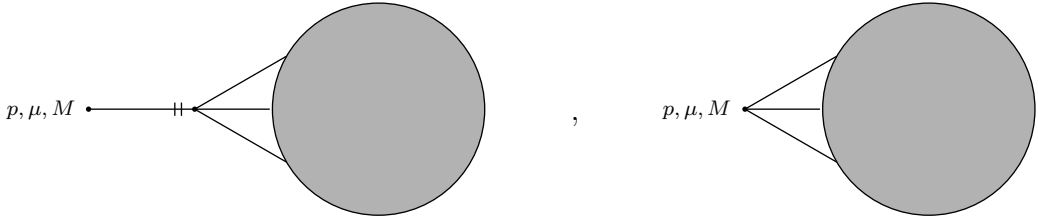
\begin{figure}
\begin{center}
\raisebox{\eqoff}{%
\begin{picture}(190,80)(0,0)\scriptsize
%\Boxc(95,40)(190,80)
\SetOffset(35,0)
\Line(0,40)(40,40)\Text(-5,40)[r]{$p,\mu,M$}
\Vertex(0,40){1}
\Line(32.5,37.5)(32.5,42.5)
\Line(35,37.5)(35,42.5)
\Vertex(40,40){1}
\Line(40,40)(74.641,60)%\Text(74.641,65)[rb]{$k_1,\alpha,A$}
\Line(40,40)(74.641,20)%\Text(74.641,15)[rt]{$k_2,\beta,B$}
\Line(40,40)(68.282,40)
\GCirc(109.282,40){40}{0.7}
\end{picture}}
$\qquad,\qquad$
\raisebox{\eqoff}{%
\begin{picture}(150,80)(0,0)\scriptsize
%\Boxc(75,40)(150,80)
\SetOffset(-5,0)
%\Line(0,40)(40,40)
%\Line(32.5,37.5)(32.5,42.5)
%\Line(35,37.5)(35,42.5)
\Vertex(40,40){1}\Text(35,40)[r]{$p,\mu,M$}
\Line(40,40)(74.641,60)%\Text(74.641,65)[rb]{$k_1,\alpha,A$}
\Line(40,40)(74.641,20)%\Text(74.641,15)[rt]{$k_2,\beta,B$}
\Line(40,40)(68.282,40)
\GCirc(109.282,40){40}{0.7}
\end{picture}}
\end{center}
\caption{Cancelling graphs from the $\propto a^3$ terms of the $\propto\theta$ terms of the SW map}
\label{cancelmech2}
\end{figure}

Altogether relying only on the $s_1$-perturbation theory we have
convinced ourselves that for $U(N)$
\be
\langle A^M_{\mu}A^N_{\nu}\rangle _{\mbox{\scriptsize kin}}~=~
-ig^2g_{\mu\nu}\delta _{AB}\frac{1}{p^2}~+~O(\theta ^3)\pnt
\label{30}
\ee
Of course, from the representation (\ref{11}) we know a priori that
there are in all orders of $\theta$ no corrections to the free propagator.
Nevertheless the above exercise was useful, since it unmasked the cancellation
mechanism for fig.~\ref{cancelmech1} and fig.~\ref{cancelmech2} as
being essential for establishing the 
already known result purely within $s_1$-perturbation theory. It is
straightforward to check also the vanishing of connected $n$-point
functions for $n>2$.\\

What changes if we switch from $U(N)$ to $G\subset U(N)$? First of all, then
we do not know the answer in advance and have to rely only on
$s_1$-perturbation theory. Secondly, in this perturbation theory the
above cancellation mechanism is set out of work for external points
carrying a primed index, related to Lie algebra elements of $U(N)$ not
in the Lie algebra of $G$. Then according to (\ref{20}) the external vertex
to start with in the first diagrams of fig.~\ref{cancelmech1} and
fig.~\ref{cancelmech2}  is zero, i.e. the partner to cancel the second
diagrams disappears. This observation is a strong hint that for
$G\subset U(N)$ there remain non-vanishing connected Green functions
$\langle A(x_1)\dots A(x_n)\rangle _{\mbox{\scriptsize kin}}$ for all
integer $n$. An explicit proof will be given in the next section. 

\section{Non-vanishing $n$-point Green functions generated by 
$\log Z_G^{\mbox{\scriptsize kin}}$} 
\label{secnptproof}

The connected Green functions 
\begin{equation*}
G_n^{\mathrm{kin},M_1\dots M_n}(x_1,\dots,x_n)=\Big\langle A^{M_1}(a(x_1))\dots A^{M_n}(a(x_n))\Big\rangle _{\mbox{\scriptsize kin} }
\end{equation*}
are power series in $\theta$ and $g$. To prove their non-vanishing for
generic $\theta$ and $g$ it is sufficient to extract at least one
non-zero contribution to $G_n^\mathrm{kin}$ of some fixed order in
$\theta$ and $g$.  

To find for our purpose the simplest tractable component of the Green
function it turns out to be   
advantageous to restrict all of the group indices $M_i$ to
primed indices that do not correspond to generators of the Lie algebra of
$G$. Then the Green function simplifies in first 
nontrivial order of the Seiberg-Witten map to:
\begin{equation}
\Big\langle\prod_{i=1}^n\big[A^{(2){m'_i}}(a(x_i))+A^{(3){m'_i}}(a(x_i))\big]\Big\rangle _{\mbox{\scriptsize kin} }\pnt
\label{fothetagreen}
\end{equation}
Here $A^{(2)}$, [$A^{(3)}$] denote the $\propto\theta$ part of the
Seiberg-Witten map  
(\ref{16}) with quadratic, [cubic] dependence on the ordinary field
$a$. Thus the above function is $\mathcal{O}\big(\theta^n\big)$.   
Focussing now on the special contribution which is exactly $\propto\theta^n$,
it is clear that in addition to the external vertices further
$\theta$-dependence (e.~g. higher order corrections to the Seiberg-Witten
map) is not allowed. That means this special part of the connected Green
function is universal with
respect to the $\theta$-expansion of the constraint (\ref{4})
where $a\in G$.

The special contribution to the Green function $\propto\theta^n$
then consists of $n$ to $\frac{3}{2}n$, [$\frac{3}{2}(n-1)+1$]
internal lines for $n$ 
even, [odd]. Two or three of these originate from each of the $n$
points (external vertices). There are no
further internal vertices present stemming from the interaction term
$s_1[a,c,\bar c ]$
in (\ref{12}) since this would increase the power in $\theta$.  

\begin{figure}
\begin{center}
\begin{picture}(330,120)(0,0)\scriptsize
%\Boxc(165,60)(330,120)
\SetOffset(110,60)
\Line(-20,34.64)(20,34.64)\Text(-25,43.3)[r]{$p_1,\mu_1,m'_1$}
\Line(-40,0)(-20,34.64)\Text(-50,0)[r]{$p_2,\mu_2,m'_2$}
\Line(-20,-34.64)(-40,0)\Text(-25,-43.3)[r]{$p_3,\mu_3,m'_3$}
\DashLine(20,-34.64)(-20,-34.64){1}%\Text(25,-43.3)[l]{$p_{n-2},\mu_{n-2},m'_{n-2}$} 
\DashLine(40,0)(20,-34.64){1}\Text(50,0)[l]{$p_{n-1},\mu_{n-1},m'_{n-1}$}
\Line(20,34.64)(40,0)\Text(25,43.3)[l]{$p_n,\mu_n,m'_n$}
\Line(-26.83,32.81)(-18.17,27.81)
\Line(-41.83,-6.83)(-33.12,-1.83)
\Line(-15,-39.64)(-15,-29.64)
\Line(26.83,-32.81)(18.17,-27.81)
\Line(41.83,6.83)(33.12,1.83)
\Line(15,39.64)(15,29.64)
\Vertex(-40,0){1}
\Vertex(-20,34.64){1}
\Vertex(20,34.64){1}
\Vertex(40,0){1}
\Vertex(20,-34.64){1}
\Vertex(-20,-34.64){1}
\Text(137.5,0)[l]{\normalsize $+\qquad\text{perm}$}
\end{picture}
\end{center}
\caption{graphs $\propto\theta^ng^{2n}$ of the connected $n$-point Green function} 
\label{diagtype}
\end{figure}
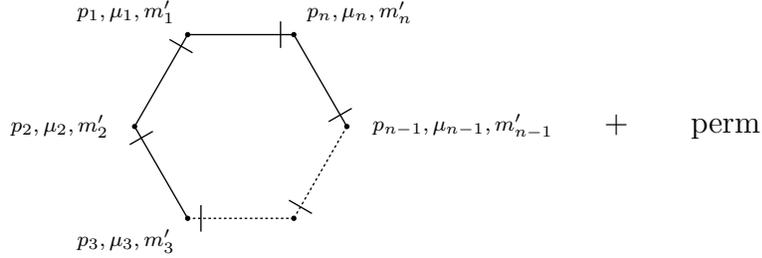

In our
normalization where the coupling constant $g$ is absorbed into the fields 
%from (\ref{funcratio}) 
each propagator enlarges the power of the
diagram in $g$ by $g^2$. Thus for general coupling $g$ it is
sufficient to check the non-vanishing of all connected diagrams with
the same number of propagators. Here we choose the minimum case of $n$
propagators where we can neglect all contributions from $A^{(3)}$ in
(\ref{fothetagreen}). Then it follows that the connected
$\propto\theta^ng^{2n}$ contributions to the Green function are
given by the type of diagrams shown in fig.~\ref{diagtype}.

The total number of the diagrams can be determined as follows: The
two lines starting at each point are distinguishable due to the
derivative at one leg. To construct all connected contributions we
connect the first leg of the first external vertex to one of the
$2n-2$ other legs that do not start at the same external point. The next
one is connected to one of the remaining $2n-4$ allowed legs, such
that no disconnected subdiagram is produced and so on. We thus have to
add-up $(2n-2)!!=(n-1)!~2^{n-1}$ diagrams. All of them can be drawn
like the one shown in fig.~\ref{diagtype} by permuting the external
momenta, Lorentz and group indices and the internal legs.    

To sum-up all diagrams it is convenient to define two classes of
permutations: 
The first includes all permutations that interchange
the two distinct 
legs at one or more external vertices
with the distribution of the 
external momenta, Lorentz and group indices held fixed. The second contains all
permutations which interchange the external quantities 
such that this cannot be traced back to a permutation of the distinct
lines at the external vertices. We call its
elements proper permutations in the following. 

In total $2^n$ combinations exist, generated by interchanging the distinct
legs when the external points are fixed. The proper permutations are
the ones which are not identical under (anti)cyclic
permutations. There are $n!$ configurations of the external points and
with each one $n-1$, [$n$] others are identified under cyclic, 
[anticyclic] permutations, i.~e. there are
$\frac{n!}{2n}=\frac{(n-1)!}{2}$ proper permutations. This is
consistent with the total number of diagrams.

The connected $\propto\theta^ng^{2n}$ contributions to the
momentum space Green function can thus be cast into the following
form:
\newlength{\arlength}
\newlength{\arheight}
\newlength{\ardepth}
\newlength{\shift}
\newcommand{\Ri}{(}
{\setlength{\fboxsep}{0pt}
\setlength{\fboxrule}{0pt}
\settowidth{\arlength}{\fbox{$\Ri$}}
\settoheight{\arheight}{\fbox{$\Ri$}}
\settodepth{\ardepth}{\fbox{$\Ri$}}
\addtolength{\arheight}{\ardepth}
\setlength{\shift}{0.5\arheight}
\addtolength{\shift}{0.2ex}
\addtolength{\shift}{-\ardepth}
\settoheight{\eqoff}{\fbox{$=$}}
\setlength{\eqoff}{0.5\eqoff}
\addtolength{\eqoff}{-60pt}
\begin{equation}
G_{n\negphantom{n{}}\phantom{\mathrm{kin},}\mu_1\dots\mu_n}^{\mathrm{kin},m'_1\dots
  m'_n}(p_1,\dots,p_n)\big|_{\propto\theta^ng^{2n}}=\sum_{\frac{\text{perm}\{i_1,\dots,i_n\}}{\text{(anti)cycl.}}}
\raisebox{\eqoff}{%
\begin{picture}(240,120)(0,0)\scriptsize
%\Boxc(120,60)(240,120)
\SetOffset(110,60)
\SetOffset(108.75,62.165)
\Vertex(-20,34.64){1}
\Line(-26.25,23.815)(-20,34.64)
\Line(-20,34.64)(-7.5,34.64)\Text(-25,43.3)[r]{$p_{i_1},\mu_{i_1},m'_{i_1}$}
\Line(-15,37.14)(-15,32.14)
\SetOffset(111.25,57.835)
\Vertex(-20,34.64){1}
\Line(-25,25.98)(-20,34.64)
\Line(-20,34.64)(-10,34.64)%\Text(-25,43.3)[r]{$p_{i_1},\mu_{i_1},m'_{i_1}$}
\Line(-24.665,31.56)(-20.335,29.06)
\SetOffset(107.5,60)
\Vertex(-40,0){1}
\Line(-33.75,-10.825)(-40,0)
\Line(-40,0)(-33.75,10.825)\Text(-50,0)[r]{$p_{i_2},\mu_{i_2},m'_{i_2}$}
\Line(-39.665,5.58)(-35.335,3.08)
\SetOffset(112.5,60)
\Vertex(-40,0){1}
\Line(-35,-8.66)(-40,0)
\Line(-40,0)(-35,8.66)%\Text(-50,0)[r]{$p_{i_2},\mu_{i_2},m'_{i_2}$}
\Line(-39.665,-5.58)(-35.335,-3.08)
\SetOffset(108.75,57.835)
\Vertex(-20,-34.64){1}
\Line(-20,-34.64)(-26.25,-23.815)\Text(-25,-43.3)[r]{$p_{i_3},\mu_{i_3},m'_{i_3}$}
\Line(-7.5,-34.64)(-20,-34.64)
\Line(-24.665,-31.56)(-20.335,-29.06)
\SetOffset(111.25,62.165)
\Vertex(-20,-34.64){1}
\Line(-20,-34.64)(-25,-25.98)%\Text(-25,-43.3)[r]{$p_{i_3},\mu_{i_3},m'_{i_3}$}
\Line(-10,-34.64)(-20,-34.64)
\Line(-15,-37.14)(-15,-32.14)
\SetOffset(111.25,57.835)
\Vertex(20,-34.64){1}
\Line(26.25,-23.815)(20,-34.64)
\Line(20,-34.64)(7.5,-34.64)%\Text(25,-43.3)[l]{$p_{i_{n-2}},\mu_{i_{n-2}},m'_{i_{n-2}}$}
\Line(15,-37.14)(15,-32.14)
\SetOffset(108.75,62.165)
\Vertex(20,-34.64){1}
\Line(25,-25.98)(20,-34.64)
\Line(20,-34.64)(10,-34.64)%\Text(25,-43.3)[l]{$p_{i_{n-2}},\mu_{i_{n-2}},m'_{i_{n-2}}$}
\Line(24.665,-31.56)(20.335,-29.06)
\SetOffset(112.5,60)
\Vertex(40,0){1}
\Line(33.75,10.825)(40,0)
\Line(40,0)(33.75,-10.825)\Text(50,0)[l]{$p_{i_{n-1}},\mu_{i_{n-1}},m'_{i_{n-1}}$}
\Line(39.665,-5.58)(35.335,-3.08)
\SetOffset(107.5,60)
\Vertex(40,0){1}
\Line(35,8.66)(40,0)
\Line(40,0)(35,-8.66)%\Text(50,0)[l]{$p_{i_{n-1}},\mu_{i_{n-1}},m'_{i_{n-1}}$}
\Line(39.665,5.58)(35.335,3.08)
\SetOffset(111.25,62.165)
\Vertex(20,34.64){1}
\Line(20,34.64)(26.25,23.815)
\Line(7.5,34.64)(20,34.64)\Text(25,43.3)[l]{$p_{i_n},\mu_{i_n},m'_{i_n}$}
\Line(24.665,31.56)(20.335,29.06)
\SetOffset(108.75,57.835)
\Vertex(20,34.64){1}
\Line(20,34.64)(25,25.98)
\Line(10,34.64)(20,34.64)%\Text(25,43.3)[l]{$p_{i_n},\mu_{i_n},m_{i_n}$}
\Line(15,37.14)(15,32.14)
\SetOffset(110,60)
\Line(-27.5,21.65)(-32.5,12.99)
\Text(-27.5,21.65)[]{\begin{rotate}{60}\raisebox{-0.5\shift}[\arheight][\arlength]{%
\hspace{-0.3\arlength}\fbox{$\Ri$}}\end{rotate}}
\Text(-32.5,12.99)[]{\begin{rotate}{-120}\raisebox{-0.5\shift}[\arheight][\arlength]{\hspace{-0.3\arlength}\fbox{$\Ri$}}\end{rotate}}
\Line(-27.5,-21.65)(-32.5,-12.99)
\Text(-27.5,-21.65)[]{\begin{rotate}{-60}\raisebox{-0.5\shift}[\arheight][\arlength]{\hspace{-0.3\arlength}\fbox{$\Ri$}}\end{rotate}}
\Text(-32.5,-12.99)[]{\begin{rotate}{120}\raisebox{-0.5\shift}[\arheight][\arlength]{\hspace{-0.3\arlength}\fbox{$\Ri$}}\end{rotate}}
\DashLine(-5,-34.64)(5,-34.64){1}
\Text(-5,-34.64)[]{\begin{rotate}{180}\raisebox{-0.5\shift}[\arheight][\arlength]{\hspace{-0.3\arlength}\fbox{$\Ri$}}\end{rotate}}
\Text(5,-34.64)[]{\begin{rotate}{0}\raisebox{-0.5\shift}[\arheight][\arlength]{\hspace{-0.3\arlength}\fbox{$\Ri$}}\end{rotate}}
\DashLine(27.5,-21.65)(32.5,-12.99){1}
\Text(27.5,-21.65)[]{\begin{rotate}{-120}\raisebox{-0.5\shift}[\arheight][\arlength]{\hspace{-0.3\arlength}\fbox{$\Ri$}}\end{rotate}}
\Text(32.5,-12.99)[]{\begin{rotate}{60}\raisebox{-0.5\shift}[\arheight][\arlength]{\hspace{-0.3\arlength}\fbox{$\Ri$}}\end{rotate}}
\Line(27.5,21.65)(32.5,12.99)
\Text(27.5,21.65)[]{\begin{rotate}{120}\raisebox{-0.5\shift}[\arheight][\arlength]{\hspace{-0.3\arlength}\fbox{$\Ri$}}\end{rotate}}
\Text(32.5,12.99)[]{\begin{rotate}{-60}\raisebox{-0.5\shift}[\arheight][\arlength]{\hspace{-0.3\arlength}\fbox{$\Ri$}}\end{rotate}}
\Line(-5,34.64)(5,34.64)
\Text(-5,34.64)[]{\begin{rotate}{180}\raisebox{-0.5\shift}[\arheight][\arlength]{\hspace{-0.3\arlength}\fbox{$\Ri$}}\end{rotate}}
\Text(5,34.64)[]{\begin{rotate}{0}\raisebox{-0.5\shift}[\arheight][\arlength]{\hspace{-0.3\arlength}\fbox{$\Ri$}}\end{rotate}}
\end{picture}}
\pnt
\label{fogreen}
\end{equation}}
Here the brackets around the external vertices denote a sum over both
configurations where the two legs are interchanged. These $n$ sums are 
then multiplied, describing exactly the $2^n$ permutations of the
distinct two legs at each vertex. 

The sum of the two permutations at one external vertex occurring $n$
times in (\ref{fogreen}) reads 
\settoheight{\eqoff}{\fbox{$=$}}
\setlength{\eqoff}{0.5\eqoff}
\addtolength{\eqoff}{-40pt}
\begin{equation*}
\raisebox{\eqoff}{%
\begin{picture}(200,80)(0,0)\scriptsize
%\Boxc(100,40)(200,80)
\Text(30,40)[r]{$p,\mu,M$}
\Vertex(40,40){1}
\Line(40,40)(74.641,60)
\Line(43.08,44.665)(45.58,40.335)
\Line(40,40)(74.641,20)
\Text(94.641,40)[]{\normalsize $+$}
\SetOffset(74.641,0)
\Vertex(40,40){1}
\Line(40,40)(74.641,60)
\Line(40,40)(74.641,20)
\Line(43.08,35.335)(45.58,39.665)
\Text(84.641,60)[l]{$q,\alpha,A$}
\Text(84.641,20)[l]{$r,\beta,B$}
\end{picture}}
~=~-\frac{i}{4}~d_{abm'}\Big[2(\theta^{\beta\gamma}q_\gamma\delta_\mu^\alpha+\theta^{\alpha\gamma}r_\gamma\delta_\mu^\beta)+\theta^{\alpha\beta}(q_\mu-r_\mu)\Big]\pnt
\end{equation*}
Using this, the analytic expression for the
$\propto\theta^ng^{2n}$ part of the connected Green function is
given by
\begin{equation}
\begin{aligned}
& G_{n\negphantom{n{}}\phantom{\mathrm{kin},}\mu_1\dots\mu_n}^{\mathrm{kin},m'_1\dots m'_n}(p_1,\dots,p_n)\big|_{\propto\theta^ng^{2n}}\\
& \qquad
=\sum_{\frac{\text{perm}\{i_1,\dots,i_n\}}{\text{(anti)cycl.}}}\frac{g^{2n}}{4^n}\int\frac{d^Dk}{(2\pi)^D}\prod_{r=1}^nd_{a_ra_{r+1}m'_{i_r}}\Big[-2\theta^{\alpha_r\gamma_r}(q_{r-1})_{\gamma_r}
g_{\mu_{i_r}\alpha_{r+1}}+2\theta_{\alpha_{r+1}}^{\phantom{\alpha_{r+1}}\gamma_r}(q_r)_{\gamma_r}
\delta_{\mu_{i_r}}^{\alpha_r}\\
&\phantom{ \qquad
=\sum_{\frac{\text{perm}\{i_1,\dots,i_n\}}{\text{(anti)cycl.}}}\frac{g^{2n}}{4^n}\int\frac{d^Dk}{(2\pi)^D}\prod_{r=1}^nd^{a_ra_{r+1}m'_{i_r}}\Big[}-\theta^{\alpha_r}_{\phantom{\alpha_r}\alpha_{r+1}}(q_{r-1}+q_r)_{\mu_{i_r}}\Big]\frac{1}{q_{r-1}^2}\col
\end{aligned}
\label{greenkinexpl}
\end{equation}
where summation over $\alpha_r$ appearing twice in the sequence of
multiplied square brackets is understood. Thereby one has to
identify $a_{n+1}=a_1$, $\alpha_{n+1}=\alpha_1$, 
$p_{i_n}=-\sum_{r=1}^{n-1}p_{i_r}$. The $q_r$ are defined by
\begin{equation}
q_r=q_r(k,p_{i_1},\dots,p_{i_r})=k+\sum_{s=1}^rp_{i_s}\pnt
\label{qdef}
\end{equation} 
In appendix~\ref{appnptproof} we prove that this expression is indeed
non-zero at least for even $n$ and the most symmetric
non-trivial configuration of the external momenta, Lorentz and group indices. 
This means that non-vanishing connected $n$ point functions for arbitrary high
$n$ exist in
the kinetic perturbation theory, leading to infinitely many building
blocks in the $\theta$-summed case. In other words one needs infinitely many
elements to formulate Feynman rules for the non-commutative $G$-gauge
theory if one insists on keeping the non-commutative $U(N)$ vertices as 
components.       

Due to the fact that the expressions discussed above cannot be
affected by higher order corrections of (\ref{4}) this statement
is universal, i.~e. independent of the power in $\theta$ to which the
constraint $a\in G$ is implemented.

\section{The case with sources restricted to the Lie algebra of $G$}
\label{secressource}
Up to now we have looked for Feynman rules working with the original
$U(N)$ vertices and sources $J^M$ taking values in the full $U(N)$
Lie algebra. This seemed to be natural since in the enveloping
algebra approach for $G\subset U(N)$ the non-commutative gauge $A^M$ field, 
although constrained, carries indices $M$ running over all generators
of $U(N)$.

There is still another option to explore. First one can restrict the sources
$J,~\eta ,~\bar{\eta}$ in (\ref{9}) by hand to take values in the Lie algebra
of $G$ only. Then instead of pulling out in (\ref{10}) the complete
interaction $S_I$ one separates only those parts of $S_I$, which yield
vertices whose external legs carry lower case Latin indices referring to the Lie
algebra of $G$ exclusively. The remaining parts of $S_I$, generating
vertices with at least one leg owning a primed index, are kept under
the functional integral. The functional integration and the constraint
remain unchanged. We denote this splitting of $S_I$ by
\be
S_I[A,C,\bar C]~=~S_i[A,C,\bar C]~+~S'_{i}[A,C,\bar C]
\label{bb1}
\ee
and the sources by hatted quantities
\be
\widehat J^{a'}~=~\widehat{\bar{\eta}}^{a'}~=~\widehat{\eta} ^{a'}~=~0~.
\label{bb2}
\ee
Then
\be
Z_G[\widehat J,\widehat{\bar{\eta}},\widehat{\eta }]~=~e^{iS_{i}[\frac{\delta}{i\delta \widehat J},\frac{\delta}{i\delta\widehat{\bar{\eta}}},\frac{\delta}{i\delta \widehat{\eta}}]}~\widehat Z_G[\widehat J,\widehat{\bar{\eta}},\widehat{\eta }]
\label{bb3}
\ee
and
\bea
\widehat Z_G[\widehat J,\widehat{\bar{\eta}},\widehat{\eta }]&=&\int _
{a,c,\bar c~\in G   }DA~D\bar C~DC~e^{i(S_{\mbox{\scriptsize kin}}[A,C,\bar C]
+S'_{i}[A,C,\bar C] + A\widehat J+\widehat{\bar{\eta}}C+\bar C\widehat{\eta )}}
\nonumber\\
&=&\int _{a,c,\bar c~\in G   } Da~D\bar c~Dc~{\cal J}~e^{i(S_{\mbox{\scriptsize kin}}[a,c,\bar c]+\widehat s_1[a,c,\bar c ]+A[a]\widehat J+\widehat{\bar{\eta}}C[c,a]+\bar c\widehat{\eta })}~,
\label{bb4}
\eea
where $\widehat s_1[a,c,\bar c ]$ is defined by
\be
S_{\mbox{\scriptsize kin}}[A[a],C[c,a],\bar c]~+~S'_{i}[A[a],C[c,a],\bar c]~=~S_{\mbox{\scriptsize kin}}[a,
c,\bar c]~+~\widehat s_1[a,c,\bar c]~.
\label{bb5}
\ee
If now 
\bea
&&\log \left (\widehat Z_G[\widehat J,\widehat{\bar{\eta}},\widehat{\eta} ]/\widehat Z_G[0]\right )~=~\nonumber\\
&&~~~~~~~~~\sum _ni^n\int dx_1\dots dx_n~\widehat J(x_1)\dots \widehat J(x_n)
\langle A(x_1)\dots A(x_n)\rangle _{\mbox{\scriptsize kin +$S'_i$}}~+~\dots ~,
\label{bb6}
\eea 
e.g. for
$G=SO(N)$, in the spirit of (\ref{13a}) would generate only the free
propagators, the $SO(N)$ Feynman rules conjectured in ref.\cite{bs2}
\footnote{There have been given arguments \cite{bsst} that their constraint
is equivalent to requiring the image under the inverse SW map to be in
$SO(N)$.}
would have been derived via partial summation of the $\theta$-expanded
perturbation theory in the enveloping algebra approach.\\

In the remaining part of this section we prove that this $cannot$ happen.
For this purpose we consider $\langle A^{m_1}(x_1)\dots
A^{m_n}(x_n)\rangle _{\mbox{\scriptsize kin +$S'_i$}}$ and look at it
as a power series in $g^2$ and $\theta $. To prove that it is not
identically zero, it is sufficient to find a particular non-vanishing
order in $g^2,~\theta $. Let us concentrate on the lowest possible
order in $g^2$. 

At all $x_i$ the contributing diagrams in the
$\widehat s_1$-perturbation theory have to start with at least one
commutative gauge field propagator (\ref{28}). This generates  at
least a factor $g^{2n}$ (One propagator at each $x_i$ corresponding to
the lowest order of SW map.) The diagrams have to be connected. To
achieve this, with respect to power counting in $g^2$, in the most
effective way one  has to connect all the $n$ legs in just one
$n$-point vertex of the $\widehat s_1$-perturbation theory, ending up
with a total $g^2$-power of $g^{2n-2}$. 

Now we search in addition for the lowest possible power
in $\theta$. 
The $n$-point vertices arise from expressing the non-commutative
fields $A$ either in the original non-commutative kinetic term or 
3-point or 4-point interactions in $S'_i$ (see (\ref{bb1})) via (\ref{16})
in terms of  the commutative field $a$.
\footnote{Note that the original 3-point or 4-point interactions by themselves
are $\theta $-dependent via the $\star$-product. But since we are searching
for lowest order in $\theta$ this further $\theta$-dependence can be
disregarded.}  
Let us look for the most efficient way for simultaneously trading a minimal 
number of  $\theta $-factors combined with a maximal number of $a$-legs. 
Simple dimensional analysis shows that this is achieved by terms in the
SW map (\ref{16}) not containing derivatives, i.e. terms of the type
$(a)^l(\theta )^{\frac{l-1}{2}}$. Then independent of the order of
contribution to the SW-map $n$-point vertices originating from the
kinetic term, the original 3-point or 4-point vertices
behave like $\theta ^{\frac{n}{2}-1},~\theta ^{\frac{n}{2}-\frac{3}{2}}$
and $ \theta ^{\frac{n}{2}-2}$, respectively. From this observation
we can conclude that for a given $n$ within the lowest $g^2$-power term
the minimal number of $\theta $ factors is exclusively realized
by connecting the $n$-external legs in just one $n$-point
vertex generated by SW-mapping out of a 4-point interaction of $S'_i$.

In appendix~\ref{appB} we prove for $SO(3)$ that the corresponding
contribution to the 8-point 
function $\langle A^{m_1}(x_1)\dots A^{m_8}(x_8)\rangle
_{\mbox{\scriptsize kin +$S'_i$}}$ is different from zero. This 
excludes the rules of \cite{bs2}. 

The more ambitious program to exclude rules based on the vertices
in $S_i$ and an arbitrary but finite number of additional building blocks
would require to show, similar to the previous section, that there 
is no $n_0$ assuring vanishing connected $n$-point functions for $n>n_0$. 
Although we have practically no doubt concerning this conjecture, a
rigorous proof is beyond our capabilities since 
for increasing $n$ higher and higher orders of the SW-map contribute.
This happens because in contrast to the proof in section~\ref{secnptproof}
one is forced to look at Green functions with all external group
indices referring to generators of the Lie algebra of $G$ since no
primed indices of the remaining generators spanning $U(N)$ are probed.  

\section{Conclusions}
Starting from the enveloping algebra approach we have studied the
issue of partial summation of $\theta$-expanded perturbation theory
for subgroups $G\subset U(N)$. The main motivation was given by the
search for some Feynman rules exhibiting UV/IR mixing similar to the
well known $U(N)$ case. The original Feynman rules in the enveloping
algebra approach contain an infinite number of vertices. They are read
off from the interactions in terms of the commutative gauge field
$a_{\mu}$ (and ghosts) taking values in the Lie algebra of $G$. Our
aim was to decide, whether by some partial summation new
rules related to the interactions of the non-commutative gauge field
$A_{\mu}$ (and ghosts) can be derived. The non-commutative fields take
values in the Lie algebra of $U(N)$, but are constrained to be related
to the commutative fields by the Seiberg-Witten map. Coming from the side of 
$\theta$-expanded perturbation theory the non-commutative fields are
composites constructed out of the commutative fields.

With our initial formula (\ref{10}) we have decided to choose the
vertices generated by the interaction term $S_{I}$ in terms of
$A_{\mu},C,\bar C$ as part of the building blocks of the wanted
Feynman rules. Then the remaining ingredients are given by
the connected Green functions related to $Z_{G}^{\mbox{\scriptsize
    kin}}$ in (\ref{11}). We found that for $G\subset U(N),~G\neq
U(N),~M<N$, the number of legs of non-vanishing connected Green
functions generated by $\log Z_{G}^{\mbox{\scriptsize kin}}$ is not
bounded from above. Therefore, there are no Feynman rules based on the
$A,C,\bar C$ vertices in $S_{I}$ and, besides perhaps suitable modified
propagators, at most a finite number of additional building blocks
with gauge field or ghost legs. 
 
As usual in the case of no go theorems one has to be very carefully
in stressing the input made. Our negative statement is bound to the 
a priori decision to work with the $A,C,\bar C$ vertices from $S_{I}$.
Of course we cannot exclude at this stage the existence of rules
exhibiting UV/IR mixing based on some clever modification of these vertices.
We also cannot exclude that the infinite set of building blocks with
gauge field and ghost legs by means of some additional auxiliary field
could be resolved into rules with only a finite number of building blocks. 

In the $SO(N)$ case Feynman rules for the fields $A,~C,~\bar C$ have
been conjectured in ref.\cite{bs2}. In these rules vertices and
propagators carry only $SO(N)$ indices. To make contact with this
situation we have modified the set up of eqs. (\ref{10}), (\ref{11}) to
(\ref{bb3}), (\ref{bb4}). This ensures that one already has the vertices
of \cite{bs2} as building blocks. A derivation of these rules would
then require that $\log \widehat Z_G$ would generate nothing beyond a
connected two point function. However, at least for $SO(3)$ we were
able to show explicitly that there is a non-vanishing connected
8-point function. \\[10mm] 
%%%%%%%%%%%%%%%%%%%%%%%%%%%
{\bf Acknowledgments}\\
We thank R. Helling, M. Salizzoni and P. Schupp for useful discussions.
The work of C.~S. is supported by DFG via the Graduiertenkolleg 271, that
of H.~D. in part by DFG via the project DO 447/3-1 and Graduiertenkolleg 271.

\appendix

\section{}
\redefinelabel{\Alph{section}}
\label{appnptproof}
To prove that the Green function in (\ref{greenkinexpl}) is
non-zero, it is sufficient 
to show that at least one contribution to this quantity with an
\emph{independent} tensor structure is non-vanishing at some configuration of
the external momenta, Lorentz and group indices. Choosing the most symmetric
non-trivial external configuration 
\begin{equation}
p_1=\dots=p_{n-1}=p\col\qquad
p_n=-(n-1)p\col\qquad\mu_1=\dots=\mu_n=\mu\col\qquad m'_1=\dots m'_n=m'
\label{exconfig}
\end{equation}
simplifies (\ref{greenkinexpl}) considerably, e.~g.
the summation over permutations of the external
quantities simply lead to a combinatorial factor.   

We first pick out all terms where -- \emph{after} performing the
integral of (\ref{greenkinexpl}) -- the tensor structure of the
$\mu_i$ 
is purely constructed with $g_{\mu_i\mu_j}$ such that $\theta^{\alpha\beta}$
does not carry an external
Lorentz index $\mu_i$. To minimize the number of
contributing terms we choose
$\theta^{\alpha\beta}(p_i)_\beta=0$.\footnote{This can be realized for
  the choice (\ref{exconfig}).} In this case the square 
brackets in (\ref{greenkinexpl})  simplify and we use the abbreviations
\begin{equation*}
2\Big[\olett{1}_r+\olett{2}_r+\olett{3}_r\Big]k=
2\Big[-\theta^{\alpha_r\gamma_r}
g_{\mu_{i_r}\alpha_{r+1}}+\theta_{\alpha_{r+1}}^{\phantom{\alpha_{r+1}}
\gamma_r}\delta_{\mu_{i_r}}^{\alpha_r}-\theta^{\alpha_r}_{\phantom{\alpha_r}
\alpha_{r+1}}\delta_{\mu_{i_r}}^{\gamma_r}\Big]k_{\gamma_r}\col
\end{equation*}
where Lorentz indices are not written explicitly. 
For the three terms inside the bracket only the following multiplications
can produce a pure $g_{\mu_i\mu_j}$-structure
\begin{equation*}
\begin{aligned}
\olett{1}_r\olett{2}_{r+1}=&\theta^{\alpha_r\gamma_r}\theta^{\gamma_{r+1}}_{\phantom{\gamma_{r+1}}\alpha_{r+2}}g_{\mu_{i_r}\mu_{i_{r+1}}}\\
\olett{2}_r\olett{1}_{r+1}=&\vphantom{\theta}^{\gamma_r}\theta\theta^{\gamma_{r+1}}\delta_{\mu_{i_r}}^{\alpha_r}g_{\mu_{i_{r+1}}\alpha_{r+2}}\\
\olett{2}_r\olett{3}_{r+1}=&\vphantom{\theta}^{\gamma_r}\theta\theta_{\alpha_{r+2}}\delta_{\mu_{i_r}}^{\alpha_r}\delta_{\mu_{i_{r+1}}}^{\gamma_{r+1}}\\
\olett{3}_r\olett{1}_{r+1}=&\vphantom{\theta}^{\alpha_r}\theta\theta^{\gamma_{r+1}}\delta_{\mu_{i_r}}^{\gamma_r}g_{\mu_{i_{r+1}}\alpha_{r+2}}\\
\olett{3}_r\olett{3}_{r+1}=&\vphantom{\theta}^{\alpha_r}\theta\theta_{\alpha_{r+2}}\delta_{\mu_{i_r}}^{\gamma_r}\delta_{\mu_{i_{r+1}}}^{\gamma_{r+1}}\col
\end{aligned}
\end{equation*}
where we have defined
$\vphantom{\theta}^\alpha\theta\theta^\gamma=\theta^{\alpha\beta}\theta_{\beta}^{\phantom{\beta}\gamma}$.    
These products are the building blocks of the complete terms
with $n$ factors, for instance like
\begin{equation*}
\underbrace{\olett{1}_1\olett{2}_2\dots\olett{1}_{k-1}\olett{2}_k\olett{3}_{k+1}\dots\olett{3}_{j+k}
\dots}_n\col
\end{equation*}
where the $\alpha_1$ index of the first factor is contracted with the
$\alpha_{n+1}$ index of the last. 
 
Further restrictions are imposed on the complete expressions: 
The total number of factors $n$ has to be
even because one cannot construct a pure $g_{\mu_i\mu_j}$ structure
with an odd number of $\mu_i$'s. In addition the number of $\olett{3}$'s in
the complete 
product of $n$ terms has to  be even as otherwise after performing the
integral in (\ref{greenkinexpl}) one  $\theta$ would
carry an index $\mu_i$ (see equations below). Then it follows that the
numbers of $\olett{1}$'s and $\olett{2}$'s have to be identical.

Using the configuration (\ref{exconfig})
the contribution of all terms with an even number $j$ of $\olett{3}$'s
and an even number $n-j$ $\olett{1}$'s and $\olett{2}$'s can now be
written as
\begin{equation}
\begin{aligned}
& G_{n\negphantom{n{}}\phantom{\mathrm{kin},c,}\mu\dots\mu}^{\mathrm{kin},m'\dots
  m'}(p,\dots,p,-(n-1)p)\big|_{\propto\theta^ng^{2n}\text{, only }
  g_{\mu\mu}}\\
& \qquad\quad
=\frac{(n-1)!}{2}\frac{g^{2n}}{2^n}\Big[\prod_{r=1}^nd_{a_ra_{r+1}m'}\Big]\\
&\phantom{\quad\qquad={}}\times\sum_{j=0,2}^n(\underbrace{\theta\dots\theta}_n)^{\gamma_{j+1}\dots\gamma_n}\underbrace{g_{\mu\mu}\dots
g_{\mu\mu}}_{n-j}\delta_\mu^{\gamma_1}\dots\delta_\mu^{\gamma_j}
\int\frac{d^Dk}{(2\pi)^D}\frac{k_{\gamma_1}\dots
k_{\gamma_n}}{q_1^2\dots q_n^2}\Big|_{\text{only special }g}
\col
\end{aligned}
\label{relstruc}
\end{equation} 
where the factor $\frac{(n-1)!}{2}$ stems from performing the
summation over all proper permutations and $q_r=k+rp$, $r\neq n$, $q_n=k$.  
To make the above expression compact we have used some further abbreviations
which we now explain.

The relevant part of the integral in the above expression is defined
as the tensor component of the integral only made out of the metric where
the metric must not possess a mixed index pair with one index from the
set $\{\gamma_1,\dots,\gamma_j\}$ and one from the set
$\{\gamma_{j+1}\dots\gamma_n\}$. It then reads
\begin{equation}
\int\frac{d^Dk}{(2\pi)^D}\frac{k_{\gamma_1}\dots
  k_{\gamma_n}}{q_1^2\dots q_n^2}\Big|_{\text{only special }g}=I_0\sum_{\scriptsize\begin{array}{c}\text{perm} \\\{i_1,\dots,i_j\}\\
  \{i_{j+1},\dots,i_n\}\end{array}}\prod_{r=1,3}^{n-1}g_{\gamma_{i_r}
\gamma_{i_{r+1}}}\col
\label{relint}
\end{equation}
where $I_0$ denotes a scalar integral which will be discussed later. 

The tensor
$(\theta\dots\theta)^{\gamma_{j+1}\dots\gamma_n}$ in
(\ref{relstruc}) is build by summing over all possibilities to
replace $\frac{n-j}{2}$ of the $n$ summation index pairs
$(\alpha_r,\alpha_r)$ in the trace
$\tr\{\theta^n\}=\theta^{\alpha_1}_{\phantom{\alpha_1}\alpha_2}\theta^{\alpha_2}_{\phantom{\alpha_2}\alpha_3}\dots\theta^{\alpha_n}_{\phantom{\alpha_n}   
\alpha_1}$ by the index pairs
$\{(\gamma_{j+1},\gamma_{j+2}),\dots,(\gamma_{n-1},\gamma_n)\}$
keeping the ordering of the $\gamma$-pairs, i.~e. the pair
$(\gamma_{j+1},\gamma_{j+2})$ is inserted at the positions with
smallest index $r$ of all replaced $\alpha_r$ and so on. All indices
$r$ of the replaced $\alpha_r$ either have to be odd or even, since otherwise
at least two substructures
$\vphantom{\theta}^{\gamma_r}\theta\dots\theta^{\gamma_{r+1}}$ would contain
  an odd number of $\theta$'s vanishing when contracted with the
  symmetric $k_{\gamma_r}k_{\gamma_{r+1}}$ in (\ref{relstruc}). Some
examples for illustration: If $j=n$ in (\ref{relstruc}) then
$(\theta\dots\theta)=\tr\{\theta^n\}$ and there is only one
contribution. If $j=n-2$ then there are $n$
possibilities\footnote{$\frac{n}{2}$ possibilities to replace
  $\alpha_r$ with odd $r$ and $\frac{n}{2}$ to replace the ones with
  even $r$.} to replace a
pair $\alpha_r$ by the pair $(\gamma_{n-1},\gamma_n)$ such that
$(\theta\dots\theta)^{\gamma_{n-1}\gamma_n}=n~\vphantom{\theta}^{\gamma_{n-1}}
\theta\dots\theta^{\gamma_n}$. For general $j\neq n$ there are
$2\binom{n/2}{j/2}$ non-vanishing possibilities to replace summation
indices by the $\gamma$-pairs.   

The contraction of the above defined 
$(\theta\dots\theta)^{\gamma_{j+1}\dots\gamma_n}$ in
(\ref{relstruc}) with the tensor
structure of the integral (\ref{relint}) leads to a sum over
products of traces  of the form $\prod_i\tr\{\theta^{2k_i}\}$,
$k_i\in\mathbb{N}$ such that 
$\sum_i 2k_i=n$. All these products of traces include the same sign
$(\sgn\tr\{\theta^2\})^\frac{n}{2}$.\footnote{This can be proven by
  using the canonical skew-diagonal form of \cite{sz}.}
Thus all summed terms in (\ref{relstruc})  
carry the same sign such that a cancellation
mechanism between different terms cannot be present. Proving
the non-vanishing of (\ref{relstruc}) therefore only requires to show
that the group structure factor and the scalar integral $I_0$ in
(\ref{relint}) are non-zero. 

For instance, the choice $m'=N^2$, where the generator $T_{N^2}$ is
given by $T_{N^2}=\frac{1}{\sqrt{2N}}\mathds{1}$ in an $U(N)$ theory, leads to
$d_{abN^2}=\sqrt{\frac{2}{N}}\delta_{ab}$. Hence, with $\dim G$ as the
dimension of the Lie algebra of $G$
\begin{equation*}
\prod_{r=1}^nd_{a_ra_{r+1}m'}\Big|_{m'=N^2}=\Big(\frac{2}{N}\Big)^\frac{n}{2}\dim
G
%\col\qquad a_{n+1}=a_1
\end{equation*}
does not vanish.

In general the integral in (\ref{relstruc}) can be decomposed in
scalar integrals like
\begin{equation*}
\int\frac{d^Dk}{(2\pi)^D}\frac{k_{\gamma_1}\dots
  k_{\gamma_n}}{q_1^2\dots
  q_n^2}=I_0\sum_{\scriptsize\begin{array}{c}\text{perm}\\
\{i_1,\dots,i_n\}\end{array}}\prod_{r=1,3}^{n-1}g_{\gamma_{i_r}\gamma_{i_{r+1}}}+\text{
  terms containing }p_{\gamma_i}
\end{equation*}
where due to the choice (\ref{exconfig}) the $q_i$
(\ref{qdef}) now only depend on $p$ such that the above tensor
structure can only be spanned by $g_{\gamma_i\gamma_j}$ and $p_{\gamma_i}$.
Notice that in (\ref{relstruc}) only one part of the total symmetric
tensor multiplying $I_0$ given in (\ref{relint}) is needed. 
In the above expression we
now choose all indices $\gamma_1=\dots=\gamma_n=\gamma$ and the momentum
$p$ such that it has a vanishing component $p_\gamma$ for the special
choice of $\gamma$. Then one finds for $I_0$
\begin{equation*}
I_0=\frac{1}{n!}\int\frac{d^Dk}{(2\pi)^D}\frac{(k_{\gamma})^n}{q_1^2\dots
  q_n^2}\col
\end{equation*}
For even $n$ this is non-vanishing since it is positive definite after
a Wick rotation. 

Thus the expression (\ref{relstruc}) in general does not vanish for
all even $n$ implying that 
at least all connected $n$-point Green functions with an even number
of external points  are
therefore present in the kinetic theory such that it produces infinitely many
building blocks in the $\theta$-summed case.

\section{}
\redefinelabel{\Alph{section}}
\label{appB}

In this appendix we give an explicit proof for the non-vanishing
of the lowest order contribution in $g^2$ and $\theta$
to $\langle A^{m_1}(x_1)\dots A^{m_8}(x_8)\rangle _{\mbox{\scriptsize
    kin +$S'_i$}}$ in the $SO(3)$ case. 

As discussed in the main text, the contribution we are after is isolated
by connecting the 8 external legs in one 8-point vertex generated
out of a 4-point interaction of the non-commutative $A$ in $S'_i$.
Via the definition of $S'_i$, at least one of the $A$ has
 to carry a primed group index. Since we look for the lowest order in
 $\theta$ we can replace the $\star $-product by the usual
 product. The interaction then has the gauge group structure 
$f_{N_1N_2K}f_{N_3n'_4K}$. Due to the subgroup property of $G$ this is zero 
if $N_j=n_j,~j=1,2,3$. Therefore we have to start with a 4-point interaction
of the $A$ where two of them carry a primed index. Then there contribute
three interaction terms 
\bea
\frac{i}{g^2}\left (f_{m_1m_2a}f_{n'_3n'_4a}~g^{\mu _1\nu _3}g^{\mu _2\nu _4}
~+~f_{n'_4m_1a'}f_{m_2n'_3a'}~g^{\mu _1\nu _3}g^{\mu _2\nu _4}
~+~f_{n'_4m_1a'}f_{n'_3m_2a'}~g^{\mu _1\mu _2}g^{\nu _3\nu _4}\right )
\nonumber\\
~\times~A^{m_1}_{\mu _1}A^{m_2}_{\mu _2}A^{m'_3}_{\nu _3}A^{m'_4}_{\nu _4}~.~~~~~~~~~~~
\label{cc1}
\eea
Now we replace $A^{m_i}_{\mu _i}$ by $a^{m_i}_{\mu _i}$ for $~i=1,2$
and $A^{n'_i}_{\nu _i},~i=3,4$ by the term with maximum number of $a$ 
within the $\theta ^1$ contribution, see (\ref{16}), and get
\bea
\frac{i}{16g^2}~\left (f_{m_1m_2a}f_{n'_3n'_4a}~g^{\mu _1\mu _5}g^{\mu _2\mu _8}
~+~f_{m_1n'_4a'}f_{m_2n'_3a'}~(g^{\mu _1\mu _2}g^{\mu _5\mu _8}~-~
g^{\mu _1\mu _5}g^{\mu _2\mu _8})\right )\nonumber\\
\times~d_{n'_3m_3e}f_{em_4m_5}~d_{n'_4m_6k}f_{km_7m_8}~\theta ^{\mu _3\mu _4}
\theta ^{\mu _6\mu _7}~a^{m_1}_{\mu _1}a^{m_2}_{\mu _2}\dots a^{m_8}_{\mu _8}
~.
\label{cc2}
\eea
With this interaction the $g^{2\cdot 8-2}~\theta ^2$ contribution
to the Fourier transform of\\ $\langle A(x_1)\dots A(x_8)\rangle
_{\mbox{\scriptsize kin +$S'_i$}}$ becomes up to the momentum
conservation factor equal to 
\begin{equation}
\begin{aligned}
M^{\mu _1\dots \mu _8}_{m_1\dots m_8}&=\frac{i}{16}g^{14}~
\sum_{\scriptsize\begin{array}{c}\text{perm}\\
\{i_1,\dots,i_8\}\end{array}}\theta ^{\mu _{i_3}\mu_{i_4}}\theta ^{\mu _{i_6}\mu_{i_7}}~
d_{n'_3m_{i_3}e}f_{em_{i_4}m_{i_5}}~d_{n'_4m_{i_6}k}f_{km_{i_7}m_{i_8}}\\
&\phantom{=\frac{i}{16}g^{14}~
\sum_{\scriptsize\begin{array}{c}\text{perm}\\
\{i_1,\dots,i_8\}\end{array}}}
\times \Big( \frac{1}{2}(g^{\mu _{i_1}\mu _{i_5}}g^{\mu _{i_2}\mu_{i_8}}-
g^{\mu _{i_2}\mu _{i_5}}g^{\mu _{i_1}\mu _{i_8}})~f_{m_{i_1}m_{i_2}a}
f_{n'_3n'_4a}\\
&\phantom{=\frac{i}{16}g^{14}~
\sum_{\scriptsize\begin{array}{c}\text{perm}\\
\{i_1,\dots,i_8\}\end{array}}\times \Big({}}
+(g^{\mu _{i_1}\mu _{i_2}}g^{\mu _{i_5}\mu_{i_8}}-
g^{\mu _{i_1}\mu _{i_5}}g^{\mu _{i_2}\mu _{i_8}})~f_{m_{i_1}n'_4a'}f_{m_{i_2}
n'_3a'}\Big)~.
\end{aligned}
\label{cc3}
\end{equation}

We will have reached the goal of this appendix if it can be shown that the 
above quantity is different from zero. Our explicit proof of
$M^{\mu _1\dots \mu _8}_{m_1\dots m_8}\neq 0$ consists in the
numerical calculation for one special choice of 
gauge group and Lorentz indices. To minimize the calculational
effort forced by taking into account all the permutations, we looked
for an index choice with a lot of symmetry with respect to the interchange
of external legs. But we also had to avoid too much symmetry not to
produce a zero result. 

If we use the standard Gell-Mann enumeration of the nine generators
of the $U(3)$ Lie algebra, see e.g. \cite{itzy}, the generators
of the $SO(3)$ subalgebra carry the indices 2,5,7. Then our special
choice for the external legs is 
\begin{tabbing}
\qquad\qquad\qquad\qquad\=leg 1\quad\=leg 2\quad\=leg 3\quad\=leg 4\quad\=
leg 5\quad\=leg 6\quad\=leg 7\quad\=leg 8\quad\\
\qquad $m_i$:\>5\>5\>2\>5\>2\>2\>5\>2\\
\qquad $\mu _i$:\>$\lambda $\>$\lambda $\>$\mu $\>$\nu $\>$\mu $\>$\mu $\>$\nu $\>
$\mu ~~~.$
\end{tabbing} 
The chosen Lorentz indices are all spacelike and have to fulfill
\bea
\mu ~\neq\nu ~,~~~~\mu ~\neq\lambda ~, ~~~~\nu\neq\lambda \nonumber\\
\theta ^{\mu\nu}~\neq ~0 ~,~~~~ \theta ^{\mu\lambda}~=~0~,~~~~
\theta ^{\nu\lambda}~=~0~.
\label{cc4}
\eea 
Taking into account the list of vanishing $d_{ABC}$ and $f_{ABC}$ for
$U(3)$ \cite{itzy} we find
\be
M^{\mu _1\dots \mu _8}_{m_1\dots m_8} \vert _{\mbox{\scriptsize special}}
~=~6i~g^{14}~\left (\theta ^{\mu\nu}\right )^2~
f_{257}^2~\left[~\left (f_{345}d_{247}~-~f_{123}d_{157}\right)^2~+~f^2_{458}d^2_{247}~\right].
\label{cc5}
\ee
All $f$ and $d$ in (\ref{cc5}) are different from zero.

%%%%%%%%%%%%%%%%%%%%%%

\end{document}